\documentclass[numsec,webpdf,contemporary,large]{oup-authoring-template}%

\graphicspath{{Fig/}}
\usepackage{rotating}
\usepackage{graphicx}
\usepackage{makecell}
\usepackage{booktabs}
\usepackage{caption}

\theoremstyle{thmstyleone}%
\theoremstyle{thmstyletwo}%
\theoremstyle{thmstylethree}%

\begin{document}

\journaltitle{Journal Title Here}
\DOI{DOI HERE}
\copyrightyear{2023}
\pubyear{2023}
\access{Advance Access Publication Date: Day Month Year}
\appnotes{Paper}

\firstpage{1}


\title[Machine Learning for Shoeprint Matching]{Improving and Evaluating Machine Learning Methods for Forensic Shoeprint Matching}

\author[1,$\dagger$]{Divij Jain\ORCID{0009-0009-0416-150X}}
\author[2,$\dagger$]{Saatvik Kher\ORCID{0009-0002-4329-9981}}
\author[3,$\dagger$]{Lena Liang\ORCID{0009-0003-3617-9719}}
\author[1,$\dagger$]{Yufeng Wu\ORCID{0009-0002-0803-1963}}
\author[4,$\dagger$]{Ashley Zheng\ORCID{0009-0003-8415-601X}}
\author[1]{Xizhen Cai}
\author[1]{Anna Plantinga}
\author[1,$\ast$]{Elizabeth Upton\ORCID{0000-0002-1219-846X}}

\authormark{Jain et al.}

\address[1]{\orgdiv{Department of Mathematics and Statistics}, \orgname{Williams College}, \orgaddress{\street{15 Hoxsey Street}, \state{Williamstown, MA USA}, \postcode{01267}}}
\address[2]{\orgdiv{Department of Mathematics and Statistics}, \orgname{Pomona College}, \state{CA, USA}}
\address[3]{\orgdiv{Department of Statistics}, \orgname{University of Chicago}, \state{Il, USA}}
\address[4]{\orgdiv{Department of Mathematics}, \orgname{Fairfield University}, \state{CT, USA}}

\corresp[$\dagger$]{Equal Contribution}{}
\corresp[$\ast$]{Corresponding author. \href{email:email-id.com}{emu1@williams.edu}}




\abstract{We propose a machine learning pipeline for forensic shoeprint pattern matching that improves on the accuracy and generalisability of existing methods. We extract 2D coordinates from shoeprint scans using edge detection and align the two shoeprints with iterative closest point (ICP). We then extract similarity metrics to quantify how well the two prints match and use these metrics to train a random forest that generates a probabilistic measurement of how likely two prints are to have originated from the same outsole. We assess the generalisability of machine learning methods trained on lab shoeprint scans to more realistic crime scene shoeprint data by evaluating the accuracy of our methods on several shoeprint scenarios: partial prints, prints with varying levels of blurriness, prints with different amounts of wear, and prints from different shoe models. We find that models trained on one type of shoeprint yield extremely high levels of accuracy when tested on shoeprint pairs of the same scenario but fail to generalise to other scenarios. We also discover that models trained on a variety of scenarios predict almost as accurately as models trained on specific scenarios.
}

\maketitle

\section{Introduction}

Footwear outsole impressions are often found at crime scenes and can serve as powerful evidence for or against an individual's presence at a crime scene. Outsole impressions are created either when materials picked up by a shoe (e.g., dirt, paint, or blood) make contact with a surface or when an imprint is left by a shoe's outsole in a substance such as mud or sand. 

There are three tiers of shoeprint characteristics with which forensic examiners determine whether or not a known shoeprint \textbf{K} from a suspect matches the questioned shoeprint \textbf{Q} found at the crime scene \citep{park2021quantifying}. \textit{Class characteristics} consist of the macro traits of a shoeprint, such as a shoe's size, brand, make, and model. \textit{Subclass characteristics} pertain to the outsole, specifically to differences in the pattern of the outsole that occur largely during manufacturing. \textit{Individual characteristics} involve the changes to the outsole pattern that occur as a result of wear and tear. These traits are considered unique to the outsole of one specific shoe and are also known as randomly acquired characteristics (RACs). 

Methods of forensic shoeprint pattern matching have historically relied largely on visual comparisons based on guidelines proposed by Scientific Working Group for Shoeprint and Tire Tread Evidence (SWGTREAD) \citep{SWGTREAD15}. A human eye can reliably determine whether two shoes' class and subclass characteristics match, but it can be difficult, even for highly trained examiners, to determine whether two prints' RACs match. Analysing RACs, however, is the most important aspect of forensic shoeprint matching since RACs are thought to be unique to specific outsoles. As such, RACs can provide strong evidence that a shoeprint belongs to a specific shoe. Additionally, examiners currently have no way to quantify the extent to which two shoeprints match. Therefore, the focus of this work is the shoeprint matching problem at the RAC level rather than at the class or subclass characteristics level.

With such high stakes as potentially connecting a crime scene print to the wrong individual, human error ought to be avoided at all costs. Based on a study conducted by \citep{Hicklin} on the consistency of forensic examiners' decisions regarding shoeprint matches, examiners vary in terms of skill level and biases. For example, some examiners tend to claim more often that shoeprints match, leading to higher-than-average false positive rates, while others err on the side of caution and tend to make more false negative decisions. In measuring repeatability, examiners repeated their own decisions only around \(60\%\) of the time. These outcomes demonstrate a concerning lack of consensus among examiners' judgements as well as an unreliability of reproducibility of their own decisions. To limit the potential influence of human subjectivity, as well as bring about the ability to provide quantitative evidence regarding how well two prints match, researchers have been developing automated methods for forensic shoeprint pattern matching \citep{park2021quantifying, park2022effect, venkatasubramanian2021quantitative}. 

To address the limitations of human judgement, prior work provides a solid foundation to complement human work with automated approaches to shoeprint matching. However, we see room for three main extensions. 

First, there is currently a trade-off between speed and accuracy for shoeprint comparison algorithms. Graph-based approaches such as maximum clique have proven to be effective in both shoeprint alignment and similarity comparison \citep{park2021quantifying, park2022effect}, but these algorithms are very computationally expensive, since maximum clique is an NP-hard problem. Computation costs inhibit the ability to process large amounts of data to generate similarity metrics for training the downstream classification model. Point cloud alignment algorithms such as iterative closest point (ICP), alternatively, compute much faster, but are more prone to error and require a good initial guess for the alignment. Since most metrics for computing similarity between two shoeprints rely on the prints being well-aligned, a few imperfect alignments can substantially bias results. Additionally, current methods use a small number of interest points for alignment, and these points are typically selected from complex computer vision algorithms such as speeded-up robust feature (SURF) \citep{SURF} and oriented FAST and Rotated Brief (ORB) \citep{ORB} \citep{park2022effect}. As such, it is challenging for forensic examiners to understand and explain why the selected points are deemed by these algorithms to be important. 

Second, there is a need for high classification accuracy without sacrificing explainability, as explainability is valuable to forensic examiners. Prior work has implemented end-to-end computer vision approaches for shoeprint matching using convolutional neural networks (CNNs) \citep{CSAFE_POC, Budka, Ma}. Even though such methods showcase impressive AUC scores, ranging from 0.96 to 0.99 across diverse datasets, their internal workings rely on black box mechanisms. Thus, there is a mounting need for methods that offer not only high precision but also high explainability. 

Third, there is a need for a unifying model for the shoeprint matching problem that can be utilised in more general settings without requiring additional data collection. Previous studies have proposed several methods for shoeprint matching at the RAC level \citep{park2021quantifying, park2022effect}. These methods have very high performance within the setting for which they were designed (i.e., high-quality, complete shoeprints), but it is unclear how generalisable the model is to settings with blurry or partial prints or unseen models of shoes. As a result, each time examiners intend to employ this method for a new crime scene, they are required to re-train the model with either data from an existing comprehensive database or new data collected specifically for the scenario of interest. The amount of labour required in the current methodology motivates us to build a unified model that can be used to classify a variety of shoeprint qualities and shoe types. 

These potential extensions to existing literature drive our research questions: How well can a unified machine learning model perform in classifying shoeprint pairs in various crime scene scenarios? How well do these methods generalise to shoeprint scenarios on which the model has not been trained? 

To begin addressing these questions, we propose three improvements to the current feature extraction pipeline. These modifications aim to make our method more explainable, more capable of extracting features with higher discriminating power between mated and non-mated pairs, and less vulnerable to noise in shoeprint images. First, we replace the use of SURF or ORB points in existing literature \citep{park2022effect} with edge detection \citep{ziou1998edge}, so as to preserve the full information from the original shoeprint image and maintain high explainability when images are converted into point clouds in the coordinate plane. Second, we use ICP to create a speed advantage over existing methods that are based on computationally expensive algorithms such as maximum clique \citep{park2021quantifying}. We also optimise the implementation of ICP to reduce the frequency of poor alignments caused by noise, which has detrimental downstream effects on similarity metric calculations and model predictions. Third, we design novel similarity metrics based on hierarchical and k-means clustering, Jaccard index, cross-correlation, and other image-based features, which improve the classification accuracy of the existing method in \cite{hana} by 0.5\%. 

Additionally, we test the generalisability of machine learning models trained only using clean prints on 12 different crime scene scenarios. The experiments show that a machine learning model trained only on full, pristine shoeprint pairs cannot reliably classify pairs of imperfect (e.g., partial, blurry, etc.) shoeprints because the distribution metrics are prone to shift significantly across these scenarios. We discover that training model based on data with both ideal and imperfect shoeprints yields far higher classification accuracy and is robust to different types of imperfect shoeprints. 

Finally, we create an open-source Python package and build an \href{https://solemate.streamlit.app/}{interactive web app} where researchers can upload shoeprints, generate predictions, and visualise the underlying machine learning pipeline. We believe that making these tools publicly available can facilitate researchers' future work and also provide forensic experts with the opportunity to assess our methods. In doing so, we hope to make the use of machine learning models more accessible and accurate in this high-stakes domain. 

The rest of the paper is organised as follows. In Section \ref{data}, we detail the datasets we use to create different scenarios and test our methodology. We describe each step of the machine learning pipeline from image processing to classification via alignment and feature extraction in Section \ref{method}. In Section \ref{results}, we present our results and their implications. Finally, we discuss the limitations of our work as well as potential future work in Section \ref{discussion}.
\section{Datasets and Research Setting} \label{data}

\begin{figure*}[!t]%
    \centering
    \includegraphics[width=\textwidth]{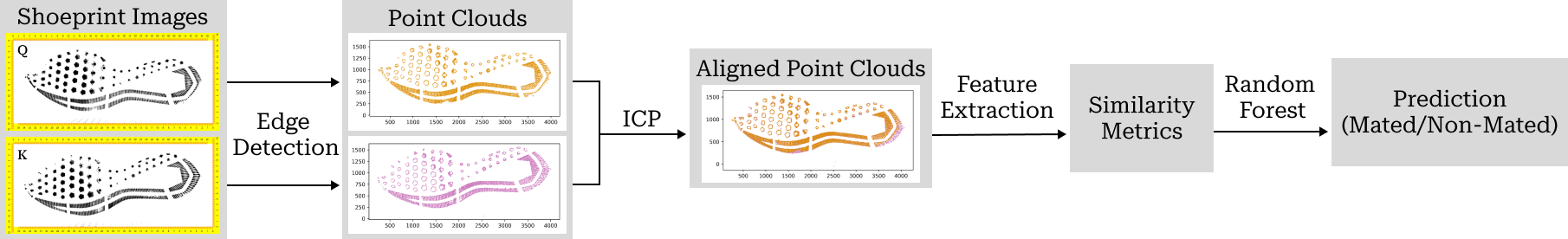}
    \caption{Flowchart of the shoeprint matching process}
    \label{fig:flowchart}
\end{figure*}

A primary goal of our research is to examine the robustness of our proposed machine learning pipeline for forensic shoeprint analysis (Figure \ref{fig:flowchart}). That is, we are interested in studying whether the similarity metrics we employ (detailed in Section \ref{metrics}) experience distribution shifts in the scenarios defined in Table \ref{tab:data_description}. These scenarios replicate a wide range of simulated crime scene shoeprint data, on which machine learning models trained on only the most ideal shoeprint scans, as well as different subsets of the outlined scenarios, will be tested. 

The data we use to train and test our machine learning pipeline were collected and maintained by the Center for Statistics and Forensic Evidence (CSAFE). We split the shoes into a training and test set, employing a 70-30 train-test split.  No shoe exists in both sets, preserving independence between the two sets.  Mated and non-mated shoeprint pairs for the training and tests sets are then created by pairing shoeprints within the training and testing sets separately.  Therefore, for each pair of shoes in the dataset, its associated shoeprints either exist in the training set or the testing set, but not both.

The first of three datasets we use is the longitudinal shoe outsole impression study \citep{park2021quantifying}. This dataset contains 2D digital scans of 158 pairs of Adidas Seeley and Nike Winflo 4 shoeprints. The shoes were worn by a group of study participants, and scans of the shoeprints were taken when the shoes were brand new and then again at three follow-up visits when the outsoles had increasing amounts of wear. At each visit, several replicate scans are taken of the same shoeprint. From this dataset, we use shoeprint scans from the first visit after the initial scan (so the prints have some amount of wear) as our baseline pristine Adidas Seeley and Nike Winflo 4 shoeprints (scenario \textbf{Pristine AN}). From these scans, we construct known mated and known non-mated pairs for the purpose of training and testing our model. For the Pristine AN scenario, we define mated pairs as different replicate scans of the same shoeprint. Constructing non-mated pairs, however, requires far more criteria. In order to best represent a realistic shoeprint comparison scenario, we want both shoe outsoles in a non-mated pair to match in all class characteristics, as ones that do not match can be easily ruled out as a mate by visual comparison. As such, we ensure that the two shoes of a non-mated pair, while originating from two different pairs and worn by different study participants, are the same model (either Adidas Seeley or Nike Winflo 4), the same size, and the same foot (either left or right). For each of the mated and non-mated pairs, we designate one shoeprint to be \textbf{Q} and the other to be \textbf{K}. 

\begin{table*}[th]
    \centering
    \footnotesize 
    \tabcolsep=4pt 
    \begin{tabularx}{\textwidth}{p{1.2cm} X X X X}

        \toprule
        \textbf{Scenario} & \textbf{Motivation} & \textbf{Data Source} & \textbf{KM Definition} & \textbf{KNM Definition} \\

        \midrule
        \RaggedRight{Pristine AN (baseline)} & 
        \RaggedRight{Pristine shoeprint scans that have been used to develop existing shoeprint matching algorithms} & 
        \RaggedRight{Pairs are constructed from visit 1 of the longitudinal dataset} & 
        \RaggedRight{Same shoe, different image replicates (657 pairs in train and 257 in test)} & 
        \RaggedRight{Different shoes, same size, same foot (L/R), same model (Nike/Adidas) (657 pairs in train and 257 in test)} \\

        \midrule
        \RaggedRight{Partial Toe, Heel, Inside, Outside} & 
        \RaggedRight{To represent a variety of incomplete shoeprints that could be recovered from a crime scene} & 
        \RaggedRight{Same pairs as baseline, \textbf{Q} is cut into a partial print} & 
        \RaggedRight{Same shoe, different image replicates (657 pairs in each of toe/heal/inside/outside train, 257 pairs in each of toe/heal/inside/outside test)} & 
        \RaggedRight{Different shoes, same size, same foot (L/R), same model (Nike/Adidas) (657 pairs in each of toe/heal/inside/outside train, 257 pairs in each of toe/heal/inside/outside test)} \\

        \midrule
        \RaggedRight{Pristine Time 2, Time 3} & 
        \RaggedRight{To simulate when an outsole undergoes wear between when a crime scene print is left and when a known scan is taken} & 
        \RaggedRight{\textbf{Q} is from visit 1 of the longitudinal dataset, 
        \textbf{K} is from visit 2 or 3} & 
        \RaggedRight{Same shoe, different image replicates, different visit number (641 pairs in each of Time 2/Time 3 train, 255 pairs in each of Time 2/Time 3 test)} & 
        \RaggedRight{Different shoes, same size, same foot (L/R), same model (Nike/Adidas), different visit number (641 pairs in each of Time 2/Time 3 train, 255 pairs in each of Time 2/Time 3 test)} \\

        \midrule
        \RaggedRight{Blurry 02, \ldots, Blurry 10} & 
        \RaggedRight{To simulate when a crime scene print is smudged or the image taken of the print is blurry} & 
        \RaggedRight{Pairs are constructed from the blurry shoeprint dataset, \textbf{K} has blur level 0, \textbf{Q} has blur level 2, 4, 6, 8, or 10} & 
        \RaggedRight{Same shoe, different blur level (306 pairs in each of Blurry 02/\dots/Blurry 10 train, 126 pairs in each of Blurry 02/\dots/Blurry 10 test)} & 
        \RaggedRight{Different shoes, same size, same foot (L/R), different blur level (306 pairs in each of Blurry 02/\dots/Blurry 10 train, 126 pairs in each of Blurry 02/\dots/Blurry 10 test)} \\

        \midrule
        \RaggedRight{Pristine 150} & 
        \RaggedRight{To test when a specific shoe make/model has not been seen by our classification model} & 
        \RaggedRight{Pairs are constructed from the 2D footwear dataset} & 
        \RaggedRight{Same shoe, different image replicates (500 pairs in train, 500 pairs in test)} & 
        \RaggedRight{Same shoe, different foot (L/R) reflected (500 pairs in train, 500 pairs in test)} \\
        \bottomrule
        
    \end{tabularx}
    \caption{Description of baseline and four simulated crime scene scenarios; KM = known mated shoeprint pairs; KNM = known non-mated shoeprint pairs}
    \label{tab:data_description}
\end{table*}

\begin{figure}[h]
  \centering
    \includegraphics[width=0.5\textwidth]{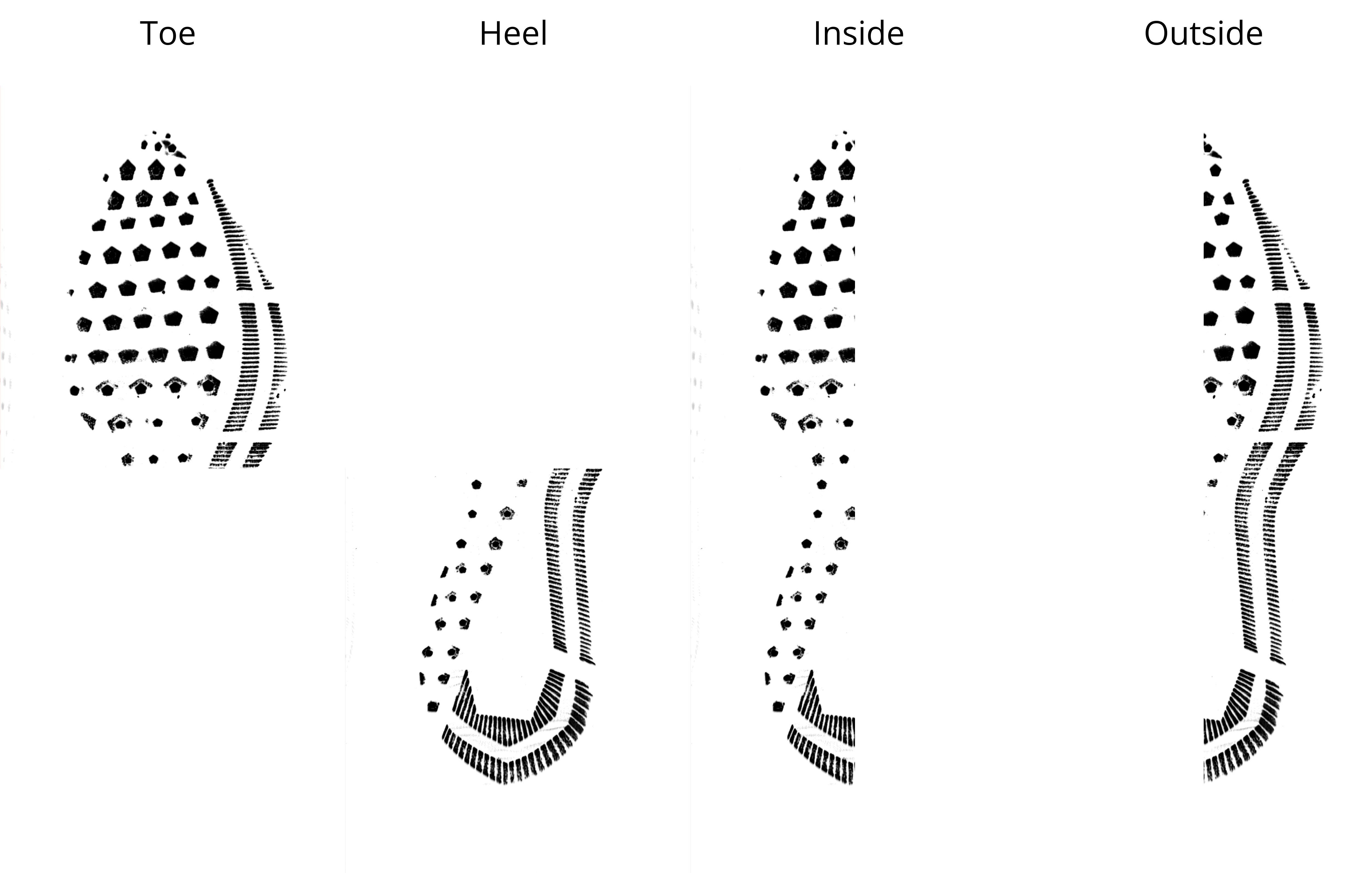}
  \caption{Simulated partial shoeprints by cutting }\label{fig:partials}
\end{figure}

The first set of scenarios to capture imperfect crime scene data is partial shoeprints. To simulate partial shoeprint data, we use the same mated and non-mated pairs as the baseline data, but for each pair, we create \textbf{Q} by manually cutting the full print. As shown in Figure \ref{fig:partials}, we conduct four different cuts: toe, heel, inside, and outside (scenarios \textbf{Partial Toe}, \textbf{Partial Heel}, \textbf{Partial Inside}, \textbf{Partial Outside}). To designate the middle $x$- and $y$-coordinate at which we make the horizontal and vertical cut, we take the arithmetic mean of the 0.025 and 0.975 quantiles of all of the $x$- and $y$-coordinates, respectively. We use these quantiles rather than averaging the minimum and maximum $x$- and $y$-values because the shoeprint scans occasionally have some noise that lies far from the rest of the shoeprint. For example, if the majority of the shoeprint spans $x$-values 0-1000, but there is some noise with $x$-values near 1500, then the middle of the shoeprint for the vertical cut would be determined to be at 750 when it ought to be closer to 500. 

We also use the longitudinal dataset to simulate the second set of scenarios: temporal shifts. The purpose of this scenario is to test whether automated algorithms can judge that a shoeprint forms a mated pair with another print even if the outsole undergoes significant wear between the two images of the shoeprint being taken. We designate \textbf{Q} as a shoeprint from visit 1 of the longitudinal dataset, and \textbf{K} as a shoeprint from visit 2 (five weeks after visit 1, scenario \textbf{Pristine Time 2}) or visit 3 (ten weeks after visit 1, scenario \textbf{Pristine Time 3}). Mated shoeprint pairs are created using shoeprints from the same individual and the same foot of varying visits while non-mated pairs are created using shoeprints from different individuals and the same foot from varying visits. The subjects made to wear the shoes were instructed to take at least 50,000 steps in the five-week period, but in most cases, they took 100,000-200,000 steps between visits. 

\begin{figure}[h]
  \centering
    \includegraphics[width=0.5\textwidth]{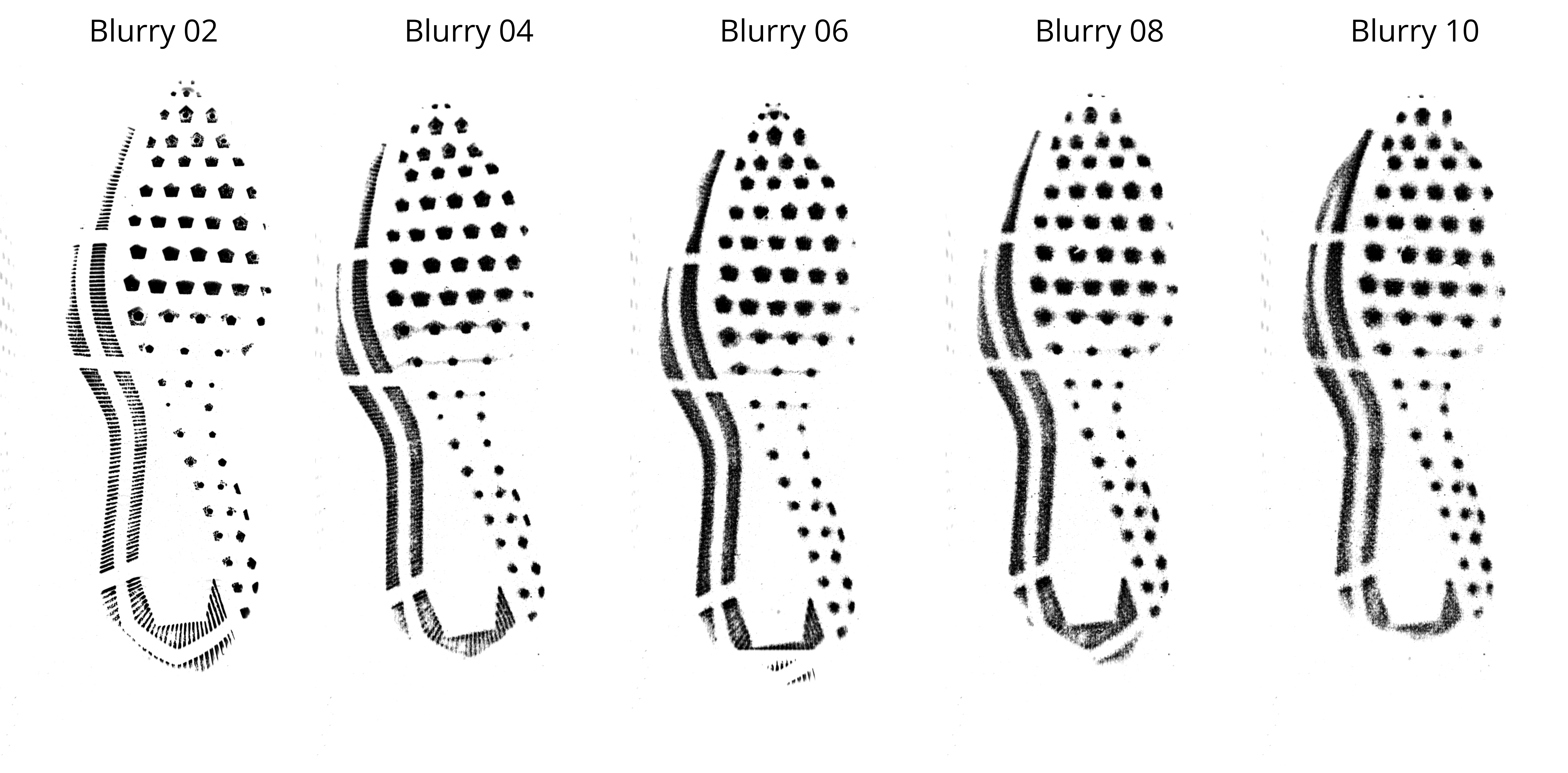}
  \caption{Increasing levels of blur, from 02 - 10}\label{fig:blurry}
\end{figure}

To test our algorithms on blurry shoeprints, we use a dataset engineered by CSAFE for this purpose \citep{park2022effect}. Using 24 pairs of Nike Winflo 4 shoes, researchers scanned the shoe outsoles first with no obstruction, then with two, four, six, eight, and ten sheets of paper between the outsole and the scanner to represent increasing levels of blurriness. We create mated shoeprint pairs of a specified blur level (2, 4, 6, 8, or 10) by matching a shoeprint with blur level 0 with a shoeprint belonging to the same individual and foot with the blur level of interest. Non-mated shoeprint pairs are then created by matching a shoeprint with blur level 0 with a shoeprint of the same foot belonging to a different individual with the same varied blur level enquired in the mated set. In all cases, we designate \textbf{K} to be blur level 0 and \textbf{Q} to be the blur level of interest (2, 4, 6, 8, or 10). We choose this as \textbf{K} is meant to represent a clean scan taken by a forensic examiner, and \textbf{Q} represents a blurry print recovered from a crime scene (scenarios \textbf{Blurry 02}, \textbf{Blurry 04}, \textbf{Blurry 06}, \textbf{Blurry 08}, \textbf{Blurry 10}, as seen in Figure \ref{fig:blurry}). 

Finally, we incorporate data from scans of outsoles of shoes of different makes and models than just Nike Winflo 4 and Adidas Seeley. CSAFE's 2D footwear dataset contains scans of 150 pairs of used shoes of a wide variety of brands and models (scenario \textbf{Pristine 150}). We once again construct mated pairs by coupling different scans of the same outsole, but there is no similar way to make non-mated pairs as there are not multiple shoes in the dataset of the same size and model. We also cannot choose shoes of different sizes and models since we want to maintain that non-mated pairs match in class characteristics. Thus, to create non-mated pairs, after choosing \textbf{Q}, we take a print of the other foot (e.g., if \textbf{Q} is a right foot, we take the left one) to be \textbf{K} and reflect it about the north/south axis of the shoeprint so that it has the same orientation as \textbf{Q}. 

\section{Method} \label{method}

\subsection{Overview} 

In order to test the performance and robustness of machine learning methods for shoeprint pattern matching, we establish an end-to-end pipeline with which researchers and examiners can go from a pair of images of outsole impressions to a probabilistic prediction of whether or not the pair is mated. The flowchart in Figure \ref{fig:flowchart} outlines the steps of the process and the algorithms used to move along the pipeline.

To see our pipeline in action, researchers can use our interactive web app \href{https://solemate.streamlit.app/}{solemate.streamlit.app}. With our open source Python package \href{https://github.com/saatvikkher/SoleMate}{SoleMate} running in the backend, the web app takes the image files for two shoeprints as an input and outputs a visualization of the prints' alignment, graphical representations of how the shoeprint pair compares to the population distributions of every similarity metric, and a posterior probability of whether the pair is mated computed by a random forest model. The following subsections provide more details on each step of the pipeline. 

\subsection{Edge Detection}

To start our pipeline, we examine two 2D scanned images \(I_{\mathbf{Q}}\) and \(I_{\mathbf{K}}\) representing the shoeprints \(\mathbf{Q}\) and \(\mathbf{K}\) respectively. These images are grayscale and are thus represented as real-valued matrices of dimension \(M \times N\).
In the matrix representation, \(I_{\mathbf{Q}}, I_{\mathbf{K}} \in \mathbb{R}^{M \times N}\). The intensities of pixels at position \((i, j)\) in images \(I_{\mathbf{Q}}\) and \(I_{\mathbf{K}}\) are indicated by the entries in \(I_{\mathbf{Q}(i, j)}\) and \(I_{\mathbf{K}(i, j)}\) respectively, for \(1 \leq i \leq M\) and \(1 \leq j \leq N\). For grayscale images, the pixel intensities vary from 0 (black) to 255 (white). 

To process an image for feature extraction, we apply edge detection. This technique identifies locations in an image where the intensity changes rapidly, effectively capturing the outlines of shoeprint outsole features \citep{edge_detect}. Once edge detection is applied, the resultant image is inverted. The inversion results in a transformation where the detected edges, initially represented as lighter shades, are transformed into darker ones against a contrasting lighter background. This allows for easier extraction of the features of interest. For the feature extraction process, we employ the Python Pillow library's \texttt{ImageFilter} module, which applies a Laplacian kernel-based edge detection \citep{Pillow}. 

Subsequent to the inversion, the highlighted features of the processed image are converted to a point cloud representation in a two-dimensional plane. The set of points is then extracted based on the darkness threshold, rendering a distinct set of coordinates where coordinate points fall at the locations of pixels with intensities below 
(darker than) the threshold. These point sets, representing the locations of the prominent features, are referred to as \(\mathcal{Q}\) for image \(I_{\mathbf{Q}}\) and \(\mathcal{K}\) for image \(I_{\mathbf{K}}\).

\subsection{Alignment}

Accurate alignment of two point clouds is a crucial step for downstream analysis, as several of our similarity metrics assume that \(\mathcal{Q}\) and \(\mathcal{K}\) are properly aligned. The data employed in our study solely utilises shoeprint images obtained for 2D scans as outlined in Section \ref{data}. Thus, we assume uniformity in the representation of different sizes across shoeprint scans. We employ the iterative closest point (ICP) algorithm, which seeks to find the best rigid body transformation (i.e., consisting of only translation and rotation) that minimises the distance between the two shoeprint point clouds \citep{besl:92}. 

In formal terms, given two point-sets \(\mathcal{Q} = \{q_1,...,q_N\}\) and \(\mathcal{K} = \{k_1,...,k_N\}\) with \(q_i, k_j \in \mathbb{R}^2\), we find the translation \(t\) and rotation \(\theta\) that minimises the sum of the squared distances,
\begin{displaymath}
    E(t,\theta) = \sum_{i, j}||q_{i} - \theta k_{j} - t||^{2}
\end{displaymath}
where the $||\cdot||$ notation represents Euclidean distance.  

One of the primary challenges with ICP is its potential to converge to local minima rather than the global minimum, especially if the initial guess for the transformation is far from the ideal solution. This can lead to suboptimal alignments. To mitigate this, we introduce a strategic method involving multiple initial translations. We employ five distinct starts. The default start involves no shift, i.e., we try to align the point cloud from its original location. The remaining four are designed to shift one of the point clouds---specifically, the non-reference cloud---in four distinct directions, namely left, right, up, and down. The magnitude of these shifts is determined by doubling the maximum range spanned by either the \(x\) (for left/right shifts) or \(y\) (for up/down) coordinates of the non-reference cloud. 

To further optimise our alignment, we employ a two-way ICP procedure. Standard ICP typically treats one point cloud as a static reference, adjusting the other for optimal alignment. In the two-way procedure, we use both \(\mathcal{Q}\) and \(\mathcal{K}\) as the reference cloud, performing the algorithm from both perspectives. This dual approach helps to surmount the limitations of a single-direction alignment. Note that, regardless of which direction yields the more accurate alignment, we always transform \(\mathcal{K}\) to match with \(\mathcal{Q}\). Even if ICP performs better when \(\mathcal{K}\) is the reference point cloud, we apply the inverse of the transformation suggested for \(\mathcal{Q}\) to \(\mathcal{K}\). We denote the transformed \(\mathcal{K}\) point cloud as \(\mathcal{K}^*\). 

Another practical challenge in aligning point clouds, particularly when dealing with tens of thousands of data points as is the case for our shoeprint point clouds, is the computational burden. To expedite the ICP algorithm's execution, we downsample each image by randomly selecting a proportion of the total number of points available before conducting ICP. This technique reduces the point cloud's density, facilitating more efficient computation while preserving the alignment's overall accuracy. While downsampling is generally advantageous throughout the alignment process, its impact is most pronounced during the preliminary iterations, where pinpoint precision is not the primary concern. Our choice of downsample rates stems from systematic exploration. We downsample the number of points in each image to 4\%, 5\%, 6\%, 20\%, and 50\% and apply ICP to each of these rates. This allows for an optimal balance between computational efficiency and alignment accuracy, as some point clouds align better with fewer points while others align better with more. 

By employing a combination of strategies, including multi-directional shifts, two-way ICP, and tailored downsampling, we have refined our alignment process, ensuring rapid computations without compromising result quality. To choose a final alignment among all these approaches, we select the alignment with the highest proportion overlap (detailed in our discussion of similarity metrics).

\subsection{Similarity Metrics} \label{metrics}

We quantify the similarity between two aligned shoeprint point clouds by calculating a number of metrics. These metrics fall into five categories: overlap metrics, minimum distances between points, metrics based on point clusters identified by clustering algorithms, phase correlation metrics, and additional image-based metrics. Additionally, we include the number of coordinate points in \(\mathcal{Q}\) and \(\mathcal{K}\) in the random forest model (discussed in Section \ref{classification}) since the interactions between the number of points in a shoeprint and the metrics help inform classification decision.

\subsubsection{Overlap}

\paragraph{Proportion Overlap}

Park and Carriquiry use the proportion of points in two shoeprints \(\mathcal{Q}\) and \(\mathcal{K}\) that overlap as a similarity metric in classifying shoeprint pairs as mated or non-mated. They identify two points as overlapping points if their distance falls within an established threshold \citep{park2021quantifying, park2022effect}. We build off of Park and Carriquiry's methods by modifying the way ``overlap'' is defined and varying the thresholds for the proportion of overlap. 

Given an arbitrary point cloud of a shoeprint, \(\mathcal{S}\), we define proportion overlap to be the number of overlapping points in \(\mathcal{S}\) divided by the total number of points in \(\mathcal{S}\). Overlapping points are the points in one shoeprint which, after alignment, are within a certain threshold of radius distance, \(d\), to at least one point in the other shoeprint. We compute proportion overlap for \(d = \{1, 2, 3, 5, 10\}\) and in both directions; that is, \(\mathcal{Q}\) proportion overlap is the proportion of points in \(\mathcal{Q}\) that overlap with points in \(\mathcal{K}^*\), and vice versa for \(\mathcal{K}\) proportion overlap. For example, a 0.60 $\mathcal{Q}$ proportion overlap for \(d=3\) would signify that 60\% of the points in $\mathcal{Q}$ are in a circle with a three-coordinate-point radius of at least one point in $\mathcal{K}^*$. We would expect mated pairs to yield higher proportion overlaps than non-mated pairs. 

\paragraph{Jaccard Index}

The Jaccard Index, which we denote as \(JI\), quantifies the similarity between two sets by measuring the proportion of shared elements to their combined elements. Formally, given the two point clouds \(\mathcal{Q}, \mathcal{K^{*}} \subseteq \mathbb{Z}^{2},\)
the Jaccard Index of 2D point clouds can be defined as: 
\[
    JI(\mathcal{Q}, \mathcal{K^{*}}) = 
    \frac{\mathcal{Q} \cap \mathcal{K^{*}}}{\mathcal{Q} \cup \mathcal{K^{*}}}
\]

When comparing point clouds, we explored the effects of different rounding granularities on the Jaccard Index calculation. Specifically, we computed three distinct versions of the Jaccard Index by rounding each point to the nearest integer, tenth, and hundredth, respectively.

\subsubsection{Minimum Distance}

Current practise involves using the median pairwise distance of the closest points in \(\mathcal{Q}\) and \(\mathcal{K}\) as a similarity metric \citep{park2021quantifying, park2022effect}. We extend this metric by considering other summary statistics of the distribution of the closest point pair distances. 

Shoeprint point clouds \(\mathcal{Q}\) and \(\mathcal{K^*}\) are composed of \((x,y)\) coordinates in \(\mathbb{R}^2\). With the aligned point clouds, we calculate the Euclidean distance between each point in \(\mathcal{Q}\) and its closest point in \(\mathcal{K}^*\). That is, let \(\mathcal{Q}_{x,y}\) be an arbitrary point in \(\mathcal{Q}\), and suppose we have a set \(P\) of points in \(\mathcal{K}^*\) that are near \(\mathcal{Q}_{x,y}\). Letting \(d_j\) be the Euclidean distance between \(\mathcal{Q}_{x,y}\) and \(j\) for any \(j \in P\), the minimum distance \(d_k\) is the distance such that \(d_k \leq d_j \forall j \in P\). We implement this calculation using the KD tree nearest neighbour algorithm. The KD tree algorithm is computationally efficient, as it groups points in \(\mathcal{K^*}\) into regions and only calculates the distances between \(\mathcal{Q}_{x,y}\) and the points in the region that contains \(\mathcal{Q}_{x,y}\), rather than all the points in \(\mathcal{K}^*\) \citep{nearest_neighbor}. From the distribution of minimum distance values for all points in \(\mathcal{Q}\), we calculate the mean and standard deviation, as well as the 10\textsuperscript{th}, 25\textsuperscript{th}, 50\textsuperscript{th}, 75\textsuperscript{th}, and 90\textsuperscript{th} percentiles. The minimum distances and the distributional spread in these distances for mated pairs should be less than those for non-mated pairs.

\subsubsection{Clustering}

We develop an approach that uses clustering methods to identify similar structures of coordinate points in \(\mathcal{Q}\) and \(\mathcal{K}\) and quantify the similarity of these structures. Our approach utilises \(k\)-means clustering with initial centroids determined via hierarchical clustering. In this setting, the \(k\)-means clustering algorithm will partition the point cloud into \(k\) distinct, non-overlapping clusters. 

For the following, we fix our choice of \(k\). The clustering pipeline proceeds as follows: 

\begin{enumerate}
    \item {
        Apply hierarchical clustering on the point cloud \(\mathcal{Q}\) using the Ward linkage function. Calculate the centroids of the \(k\) clusters obtained from the hierarchical clustering.
    }
    \item {
        Apply the \(k\)-means algorithm to \(\mathcal{Q}\) with the calculated centroids from step 1 as the starting points. Calculate the centroids of the \(k\) clusters obtained from the \(k\)-means clustering on \(\mathcal{Q}\).
    }
    \item {
        Apply the \(k\)-means algorithm to \(\mathcal{K}^*\) with the calculated centroids from step 2 as the starting points.
    }
    \item {
        Compute similarity metrics to quantify the relationship between the clustering structures of \(\mathcal{Q}\) and \(\mathcal{K}^*\).
    }
\end{enumerate}

Suppose we obtain the clusters \(C^{\mathcal{Q}}_{1}, C^{\mathcal{Q}}_{2}, \ldots, C^{\mathcal{Q}}_{k},\) and  \(C^{\mathcal{K}^*}_{1}, C^{\mathcal{K}^*}_{2}, \ldots, C^{\mathcal{K}^*}_{k}\) from running the proposed pipeline on \(\mathcal{Q}\) and \(\mathcal{K}^*\). Each cluster in $\mathcal{K^*}$ is matched to a corresponding cluster in $\mathcal{Q}$ based on similar centroid locations (i.e. the centroids do not shift much in step 3 above).  Then, we compute four similarity metrics as follows:

\begin{itemize}
    \item {
        Centroid Distance Metric (CDM):

        We compute the root mean squared distance across corresponding pairs of cluster centroids as follows:
        
        \[\text{CDM}_{k}(\mathcal{Q}, \mathcal{K}^*) = \sqrt{\frac{1}{k}\sum_{i=1}^{k} \left|\left|\text{centroid}\left(C_{i}^{\mathcal{Q}}\right) - \text{centroid}\left(C_{i}^{\mathcal{K}^*}\right)\right|\right|^{2}}.\]
    }
    \item {
        Cluster Proportion Metric (CPM):

        For each corresponding cluster \(i\) in \(\mathcal{Q}\) and \(\mathcal{K}^*\), we compute the number of points in each cluster as a proportion of the total number of points in the entire shoeprint, then take the root mean square error of the difference in proportions for each pair of corresponding clusters:
        \[
            \text{CPM}_{k}(\mathcal{Q}, \mathcal{K}^*) = \sqrt{\frac{1}{k}\sum_{i=1}^{k}\left(\frac{|C_{i}^{\mathcal{Q}}|}{|\mathcal{Q}|} - \frac{|C_{i}^{\mathcal{K}^*}|}{|\mathcal{K}^*|}\right)^{2}}.
        \] 
     }
     \item {
        Iterations to Converge Metric (IM): 

        We denote \(\text{IM}_{k}\) as the number of iterations taken for the \(k\)-means algorithm to converge for \(\mathcal{K}^*\) when initialised with the centroids obtained from running the \(k\)-means algorithm on \(\mathcal{Q}\). 
     }
     \item {
        Total Within-cluster Variation Ratio Metric (TWRM):

        For each cluster, the within-cluster variation is defined as the mean squared distance from the points that make up the cluster to the centroid of the cluster. We first calculate the total within-cluster variation of \(\mathcal{Q}\) and \(\mathcal{K}^*\).  We normalise it according to the total number of points in the corresponding shoeprint to decrease the relative influence of smaller clusters. Thus, for a given shoeprint \(\mathcal{S}\), we have the total within-cluster variation:
        \[\begin{aligned} 
            \text{TW}_{k}(\mathcal{S}) = \frac{1}{|\mathcal{S}|}\sum_{i=1}^{k}\left(\frac{1}{|C_{i}|}\sum_{x \in C_{i}}\left|\left|x - \text{centroid}\left(C_{i}\right)\right|\right|^2\right).
        \end{aligned}\]
        We then take the ratio of the within-cluster variations of shoeprints \(\mathcal{Q}\) and \(\mathcal{K}^*\) and define the corresponding metric as follows:
        \[\text{TWRM}_{k}(\mathcal{Q}, \mathcal{K}^*) = \frac{\text{TW}_{k}(\mathcal{Q}) - \text{TW}_{k}(\mathcal{K}^*)}{\text{TW}_{k}(\mathcal{Q})}.\]
     }
\end{itemize}

We rigorously explore the optimal choice of \(k\) for our clustering methodology. Initially, we evaluate the within-cluster variation for values of \(k\) ranging from 10 to 500, with increments of 10, on a set of 100 randomly chosen shoe prints. By examining the plot of within-cluster variation against the number of clusters, we observe that \(k = 20\) serves as a significant inflection point. Thus, we select \(k = 20\) for subsequent analysis. Furthermore, we retain \(k = 100\) in our experiments because the within-cluster variation appears to stabilise and converge beyond this value as seen in Appendix \ref{app:Method}, Figure \ref{fig:cluster_elbow}. This leaves us with two sets of metrics for \(k = 20\) and \(k = 100\).

\begin{figure*}[htb!]
  \centering
    \includegraphics[width=\textwidth]{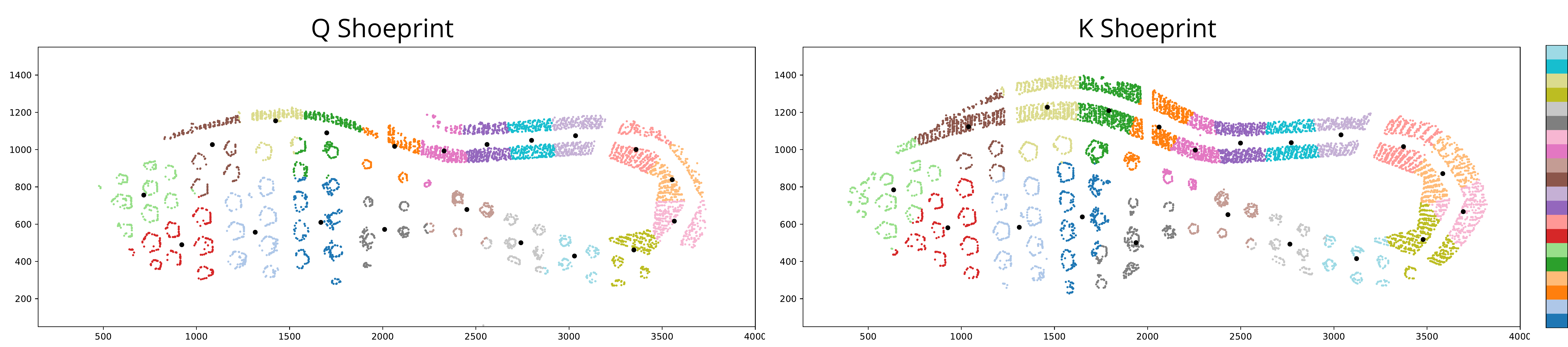}
  \caption{k-means clustering on Known-Mated Q and K shoeprints for \(k=20\) as selected by minimising WCV (Appendix \ref{app:Method}, Figure \ref{fig:cluster_elbow})}\label{fig:q_k_cluster}
\end{figure*}

The motivation for incorporating these clustering metrics lies in the expectation that a mated pair \textbf{Q} and \textbf{K} will yield similarly structured clusters after \(k\)-means clustering. For example, the distances between cluster centroids should be smaller for mated pairs than for non-mated pairs, since for mated pairs, the set of clusters in \(\mathcal{K}^*\) should be located similarly to those in \(\mathcal{Q}\).

\subsubsection{Phase Correlation}

The Fourier transform is a mathematical tool that can be utilised for signal processing and image analysis. Moreover, recent work in forensic shoeprint analysis has utilised phase correlation as a tool to align shoeprint images and construct comparative measures to identify mated and non-mated shoeprint pairs \citep{Gueham, CSAFE_POC, Gautham}. We perform phase correlation analysis using the full shoeprint image prior to edge detection. The transformation matrix calculated during ICP alignment is saved and used to align the full coordinates. In our implementation of phase correlation, we follow the algorithm first implemented by Kuglin and Hines \citep{Kuglin_Hines_1975}. 

Let \(J_{\mathbf{Q}}, J_\textbf{K} \in \mathbb{R}^{M\times N} \) be the full initial shoeprint images of Q and K, respectively. \(J_{\mathbf{Q}}\) and \(J_{\mathbf{K}}\) consist of entries in the set \(\{0, 1\}\), where 0 represents a black pixel, and 1 represents a white pixel. Due to consistent illumination across the 2D scanned images in our datasets, we designate pixels with an intensity less than a grayscale threshold of 85 (on the 0-255 scale) as black pixels. 

We begin phase correlation by applying the fast Fourier transform (FFT) algorithm \citep{FFT} to the two shoeprint images of interest, \(J_\textbf{Q}\) and \(J_\textbf{K}\). Let \( \mathcal{F}(J_{\mathbf{Q}})\) and \( \mathcal{F}(J_{\mathbf{K}})\) denote the Fourier transform of shoeprints \textbf{Q} and \textbf{K} respectively. The cross-power spectrum is then computed by multiplying the complex conjugate of \(\mathcal{F}(J_\textbf{Q})\) with \(\mathcal{F}(J_\textbf{K})\), where ``\(\circ\)'' represents the entry-wise (Hadamard) product:
\[ R(J_{\mathbf{Q}}, J_{\mathbf{K}}) = \overline{\mathcal{F}(J_{\mathbf{Q}})} \circ \mathcal{F}(J_{\mathbf{K}}) \]

We choose to omit the normalisation step in the computation of the cross-power spectrum due to the uniform illumination of our datasets. The inverse FFT is applied to the cross-power spectrum, converting it into a phase correlation map, \(r\), as an image object: 
\[r = \mathcal{F}^{-1}(R(J_{\mathbf{Q}}, J_{\mathbf{K}}))\]

The phase correlation map created by the PC process provides information on both alignment and similarity. Using the phase correlation map, we implement two similarity metrics as follows: 

\begin{itemize}
    \item Peak Value: 

    Peak value, \(PV\), corresponds to the maximum value of the phase correlation matrix. A larger peak value corresponds to stronger signal power between a shoeprint pair, so the peak value for mated shoeprints should be greater than the peak value for non-mated \citep{Foroosh_peak_val, Gueham}. Recall \(J_\mathcal{Q}\) and \(J_\mathcal{K}\) have dimensions \(M\times N\) with \(i\), \(j\) representing the elements in \(M\) and \(N\) respectively. Peak value thereby divides the maximum value of \(\mathcal{r}\) by the normalised \(\mathcal{r}\). We formally express this by: 
    \[PV = \frac{\text{max}(r)}{\frac{1}{MN}\sum_{i,j}r_{i,j}}\]
    
    \item Peak-to-Sidelobe Ratio (PSR):

    PSR reflects the relative strength of the phase correlation peak with its sidelobe levels \citep{psr}. Peak strength of phase correlation is the amplitude of the maximum peak in the phase correlation matrix. The sidelobe level denotes the average of the phase correlation map. Sidelobes express the background features and noise of an image. PSR then quantifies the strength of the image alignment, where higher values imply stronger phase correlation with less noise. We expect mated shoeprints to have higher PSR values.

\end{itemize}

\subsubsection{Additional Image-Based Metrics}

We use more image-based metrics to evaluate similarities between shoeprint images prior to edge detection. This process preserves all points in the shoeprint as well as noise captured in the original image. We implement three additional image-based metrics as described below: 

\begin{itemize}
    \item Normalised Cross-Correlation Coefficient (NCC) \citep{CSAFE_POC, Gautham}:

    NCC is calculated using the Pearson correlation coefficient of two normalised, aligned shoeprint images. It indicates the correlation of pixel intensities between the normalised images of \(\mathcal{Q}\) and \(\mathcal{K}^*\). Consider an arbitrary shoeprint \(J\); its normalised shoeprint is then \( \bar{J} = \frac{1}{MN} \sum_{i,j}{J_{(i,j)}} \). Let NCC be \(\hat{\rho} \in \mathbb{R}\) such that \(-1 \le \hat{\rho} \le 1\). The computation for NCC is as follows:
    
    \[ \hat{\rho} = \frac{\sum_{i,j}(J_{\mathcal{Q}(i,j)} - \bar{J}_{\mathcal{Q}}) \cdot (J_{\mathcal{K}^{*}(i,j)} - \bar{J}_{\mathcal{K}^{*}(i,j)})}{\sqrt{\sum_{i,j}(J_{\mathcal{Q}(i,j)} - \bar{J}_{\mathcal{Q}})^2} \cdot \sqrt{\sum_{i,j}(J_{\mathcal{K}^{*}(i,j)} - \bar{J}_{\mathcal{K}^{*}})^2}} \]

     A value of 1 indicates a perfect positive linear relationship between pixel intensities. That is, as the pixel intensity in one image increases, the pixel intensity in the other image also increases for corresponding pixel locations. Conversely, a \(\hat{\rho}=-1\) indicates a perfect negative linear relationship. With image normalisation, NCC is therefore invariant to changes in mean and scale of pixel intensities. We expect mated shoeprint pairs to be more correlated than non-mated pairs.

    \item Mean Squared Error of Images (MSE):

    MSE takes the mean squared error of the two image objects, treating each image as a \(M \times N\) matrix of pixel locations with a 0 for a black pixel and a 1 for a white pixel. For mated shoeprint pairs, we expect their MSE to be lower than those of non-mated shoeprint pairs as more of their pixels should align. We formally define MSE as: 

\[ MSE = \frac{1}{MN} \sum_{i,j}(J_{\mathcal{Q}(i,j)} - J_{\mathcal{K}(i,j)})^2 \]
    
    \item Structural Similarity Index (SSIM):

    SSIM is a measure of differences in local pixel intensities. SSIM iterates through non-overlapping squares within each image and develops a metric through the luminance and contrast within each area \citep{ssim}. We implement SSIM through the Python package \texttt{skimage.metrics} and expect SSIM to be greater for mated pairs than non-mated.
    
\end{itemize} 

\subsection{Classification} \label{classification}

After extracting a total of 35 similarity features, we construct a random forest classifier, which takes similarity features as inputs and generates an output value ranging from 0 to 1, representing the predicted posterior probability that a given pair of shoeprints is either mated (1) or non-mated (0). Explainability is important in the forensic setting because our methods must provide probative evidence that is admissible in court and convincing to a jury/judge. Hence, we choose a random forest model because it allows us to derive impurity-based feature importance for the similarity metrics. This is preferred to other ``black-box'' models such as neural networks. Additionally, unlike parametric models, a random forest is preferred because it makes few assumptions about the nature of the relationship between variables.

To build the random forest, we first partition the baseline dataset (Pristine AN) into a 70-30 train-test split. To guarantee independence between training and testing data, the partition ensures that training and testing data do not share images from the same shoe. Next, we use a five-fold cross-validation approach to fine-tune the model's parameters by grid searching through 144 different parameter combinations, including options for the number of trees (500, 1000, 2000, 5000), the maximum tree depth (10, 30, 50, unlimited), the minimum number of samples needed to split an inner node (2, 5, 10), and the minimum samples needed at a leaf node (1, 2, 4). The model with the best average cross-validation accuracy is then selected as the final model.

\section{Results} \label{results}

\subsection{Baseline Model}

Our research objective is to evaluate the feasibility of creating a single classification model that can be used for a variety of shoeprint qualities and crime-scene scenarios. Thus, we begin by creating a Baseline model using only the Pristine AN dataset and test its efficacy across the following 12 scenarios: Pristine 150, Pristine Time 2/Time 3, Blurry 02/04/06/08/10, and Partial Toe/Heel/Inside/Outside, all of which are defined in Section \ref{data}. 

Before testing the model's performance, we also split the data in these 12 scenarios into training and testing sets. Although the Baseline model is not trained with any of the data from the 12 scenarios, we split these data in order to later develop scenario-specific models as benchmarks. Table \ref{tab:train-test-all-scenarios} summarises the number of known mated (KM) and known non-mated (KNM) pairs we have for each scenario. We use a 70-30 split for all scenarios except for Pristine 150 so as to preserve a larger variety of different shoe models from Pristine 150 in the testing set. Using a 50-50 split on Pristine 150 allows us to perform thorough testing on whether a model trained only on the two shoe models from Pristine AN (Nike Winflo 4 and Adidas Seeley) can be generalised to the wider variety of shoe models found in Pristine 150. 

\begin{table}
    \centering
    \begin{tabular}{lcccc}
        \toprule
        \multirow{2}{*}{\textbf{Scenario}} & \multicolumn{2}{c}{\textbf{Train}} & \multicolumn{2}{c}{\textbf{Test}} \\
        \cmidrule(lr){2-3} \cmidrule(lr){4-5}
        & KM & KNM & KM & KNM \\
        \midrule
        Pristine AN (baseline) & 657 & 657 & 257 & 257 \\
        Pristine 150 & 500 & 500 & 500 & 500 \\
        Pristine Time 2 & 641 & 641 & 255 & 255 \\
        Pristine Time 3 & 676 & 676 & 247 & 247 \\
        Blurry 02, \ldots, Blurry 10 & 306 & 306 & 126 & 126 \\
        Partial Toe, \ldots, Outside & 657 & 657 & 257 & 257 \\
        \bottomrule
    \end{tabular}
    \caption{Number of shoeprint pairs in each scenario}
    \label{tab:train-test-all-scenarios}
\end{table}

Using the procedure described in Section \ref{classification}, we fine-tune a Pristine AN (Baseline) model using the training split in Pristine AN. Our setup achieves the highest average cross-validation accuracy of 94.83\% during grid search. We use 1000 trees with unlimited maximum tree depth, a minimum of two samples needed to split an inner node, and a minimum of one sample needed at a leaf node.

Once we train the Pristine AN model, we find that the five most important features for informing the classification decision of the random forest are NCC, Proportion Overlap K (2 px), Jaccard Index (rounded to the nearest integer), Proportion Overlap K (3 px), and Proportion Overlap K (1 px). The full variable importance plot can be found in Appendix \ref{app:Results}, Figure \ref{fig:baseline_var_imp}. 

\begin{figure}[htb]
  \centering
    \includegraphics[width=0.5\textwidth]{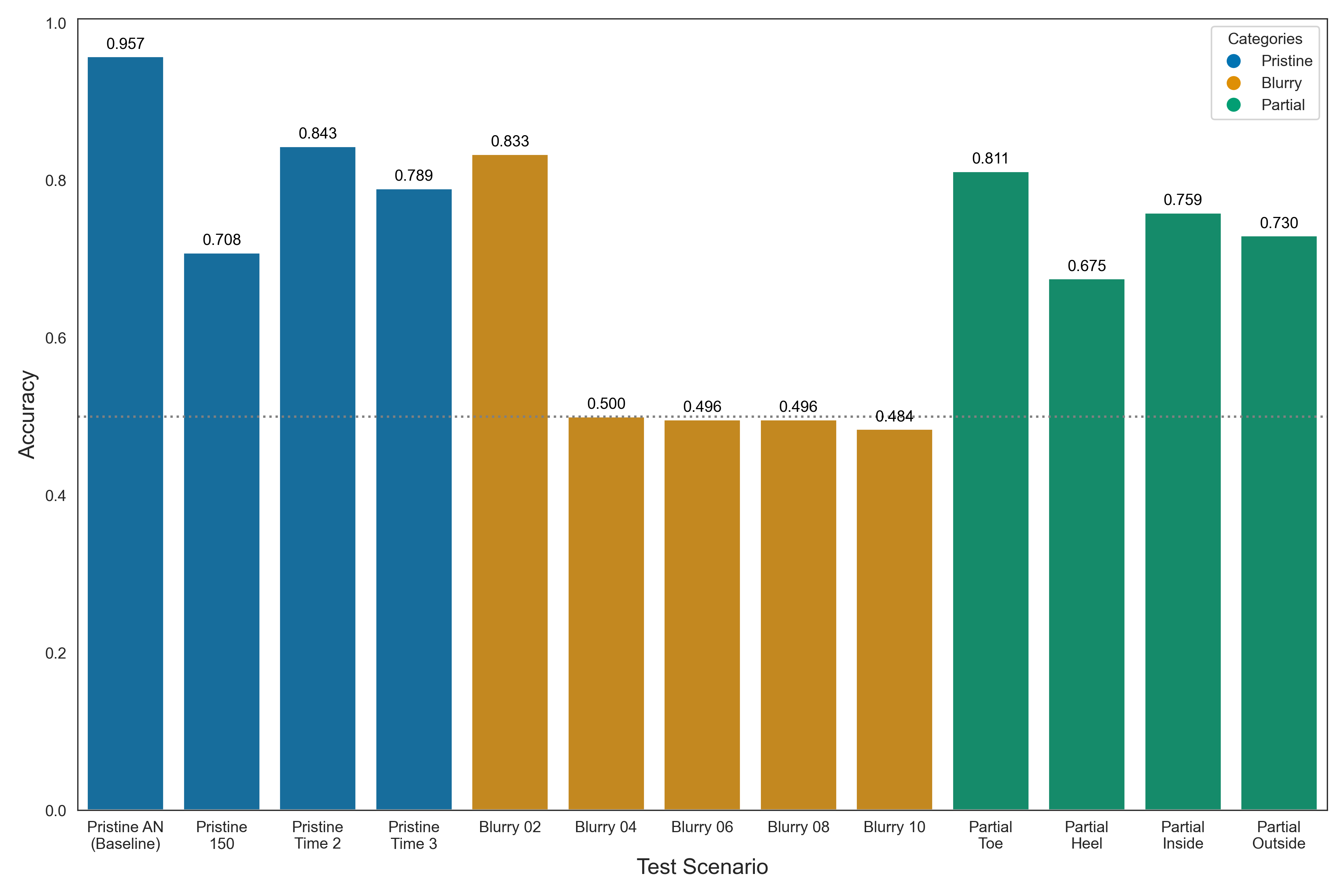}
    
  \caption{Test Accuracy (classification threshold = 0.5) of the baseline model trained on Pristine AN across all scenarios. The dashed line indicates 50\% accuracy.}
  \label{fig:baseline_test}
\end{figure}

Next, we test the Baseline model's performance on 13 shoeprint scenarios. As seen in Figure \ref{fig:baseline_test}, the model performs well on the Pristine AN test dataset (with an accuracy of 95.7\%). This result is unsurprising as the model had high cross-validation accuracy. However, the accuracy drops for partial prints, blurry prints, and pristine scans of different shoe models. These results suggest that a model trained on pristine scans of only two distinct shoe models cannot be generalised to new scenarios. 

\begin{figure*}[t]
  \centering
    \includegraphics[width=\textwidth]{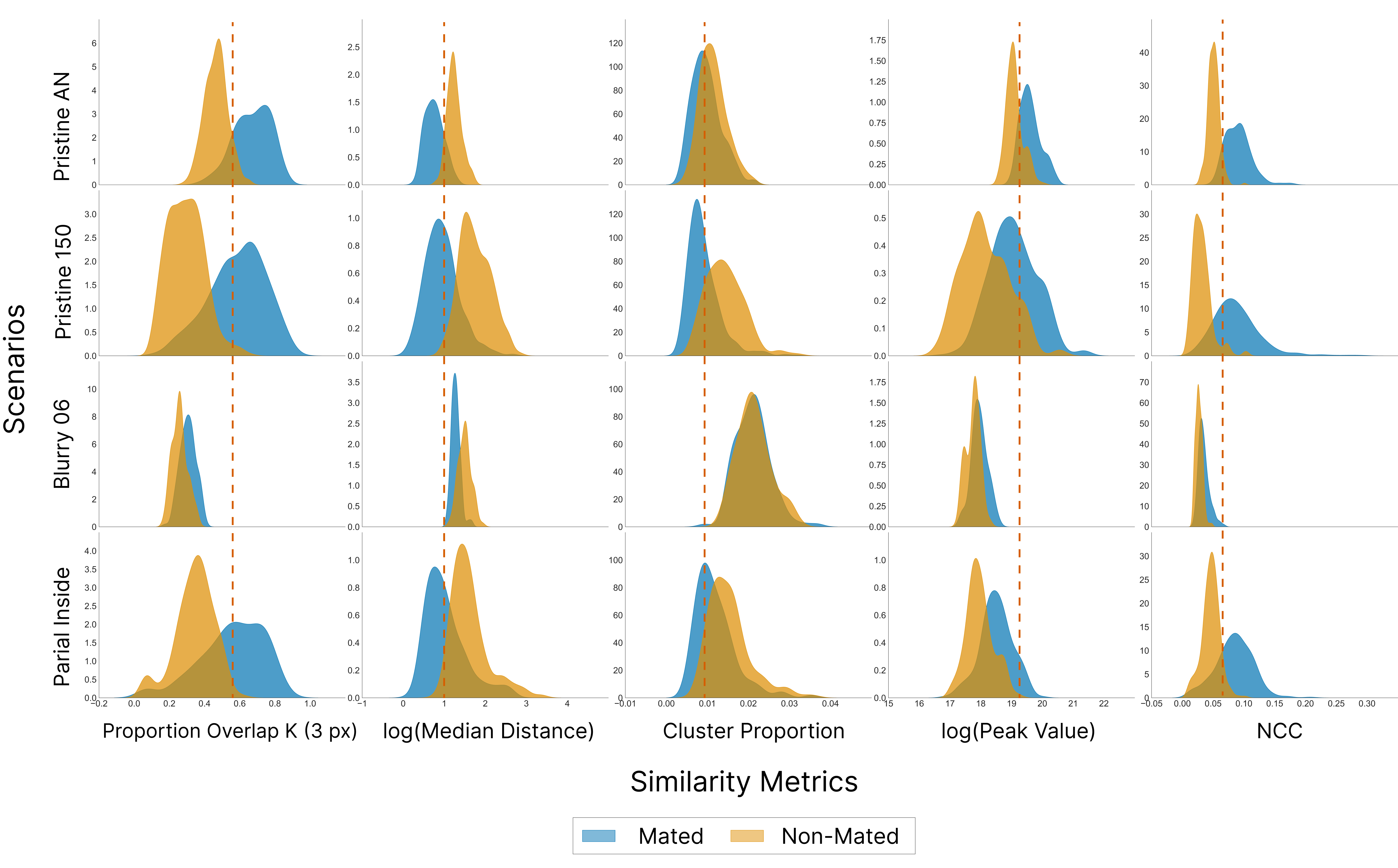}
  \caption{Distribution shifts for select metrics across scenarios. Distributions are calculated from the test set of each scenario. Kernal Density Estimation (Gaussian) is used to smooth histograms into distributions. Median Distance and Peak Value are log-transformed for better visualisation; these metrics are not log-transformed when training the random forest.}
  \label{fig:dist_shift}
\end{figure*}

Figure \ref{fig:dist_shift} provides insights into why the Baseline model performs poorly in these different scenarios, especially on blurry prints. This figure demonstrates the distribution shift phenomenon for similarity metrics across a subset of scenarios. A model trained on the Pristine AN scenario learns to classify shoeprint pairs as mated or non-mated based on the distribution of similarity metrics of this particular dataset. However, proportion overlap \(\mathcal{K}\) (with $d=3$), for example, shifts significantly to the left in the Blurry 06 scenario (first column, third row of Figure \ref{fig:dist_shift}). The Baseline model fails completely for Blurry 06 because it incorrectly classifies almost every pair as non-mated, and thus we see a testing accuracy of around 50\% for Blurry 06 in Figure \ref{fig:baseline_test}. Similarity metrics from other scenarios such as Pristine 150 and Partial Inside do not shift as drastically as those for Blurry 06, but the shifts still result in significant drops in classification accuracy, as shown in Figure \ref{fig:baseline_test}. 

An important insight from this experiment is that a model trained on one shoeprint scenario cannot generalise to other scenarios due to distribution shifts in similarity metrics. Therefore, we need to consider alternative approaches to train our random forest in order to build a unified model that is robust to multiple scenarios. 

We quantify the distribution shifts for every feature using Earth Mover's Distance, which is chosen because it is a symmetric metric. The results are displayed in Appendix \ref{app:Tables},  Table \ref{tab:earth_movers_distance_mated} and Table \ref{tab:earth_movers_distance_non_mated}. Here, larger numbers signify greater distribution shifts that likely contribute to a decrease in prediction accuracy.

\subsection{The Proposed Full Model}
\subsubsection{A. Full vs Baseline Model}

Building upon our insights from the previous section, in which we found that the baseline model is susceptible to distribution shifts, we construct a Full model to compare against the Baseline model. Our Full model incorporates the training subsets of every scenario. Furthermore, we construct two versions of the Full model: one with indicator variables for the categories (blurry, partial, and pristine; categories are explained in more detail in Section \ref{B}) and one omitting indicator variables. We aim to determine whether a random forest can inherently recognise and adjust to interactions on its own, or if it benefits from explicit human input through the use of indicator variables. 

\begin{figure*}[t]
  \centering
    \includegraphics[width=\textwidth]{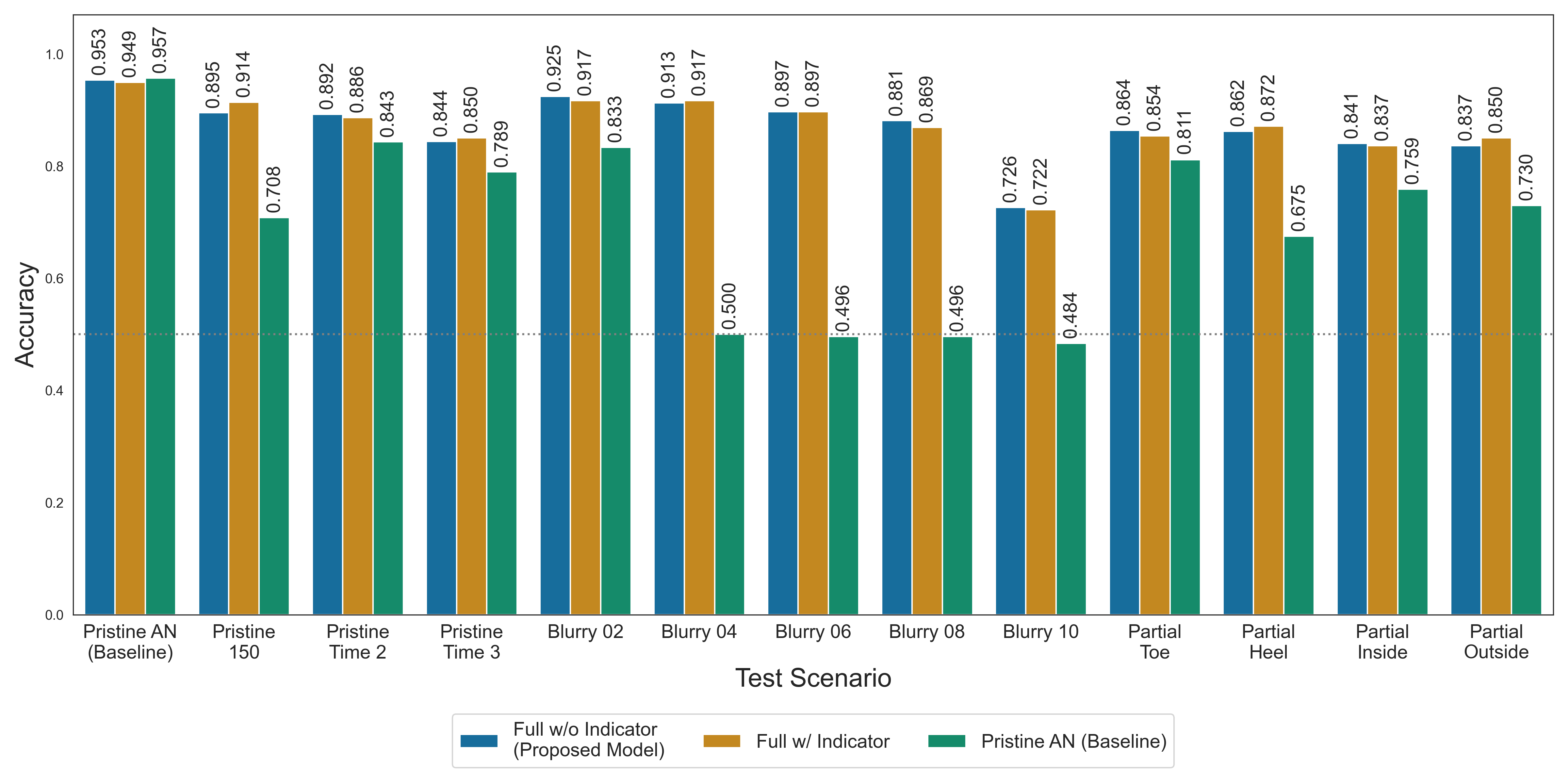}
  \caption{Comparing the test accuracy of the Full models to the Pristine AN model. The Full models show significant improvements in test accuracy.}
  \label{fig:baseline_vs_full}
\end{figure*}

As illustrated in Figure \ref{fig:baseline_vs_full}, we can observe that the Baseline model's performance is the lowest in nearly all scenarios, with the exception of the Pristine AN scenario. Notably, within this Pristine AN scenario, the performance gap between the Baseline model---which emerges as the top performer—--and the Full model with indicator variables—--the least effective for this scenario---is a marginal difference of just 0.8 percentage points. 

When examining a broader spectrum of scenarios, the distinction in performance between the Baseline model and the superior Full model is evident. The difference in accuracy spans from 4.9 percentage points to a substantial 41.7 percentage points. The disparity becomes particularly pronounced within blurry scenarios, especially levels 4-10. In these contexts, the Baseline model's accuracy exhibits a steep decline, dropping rapidly from 0.8333 at blurry level 2 to 0.4841 at blurry level 10. In contrast, the Full model maintains a robust performance, with accuracies between 0.9246 and 0.881 for blurry levels 2-8, though the model accuracy drops to 0.726 for blurry level 10. These patterns underscore the Full model's superior capacity for generalisation in comparison with the Baseline model. 

For the temporal scenarios, Pristine Time 2 and Pristine Time 3, both Full models outperform the baseline with accuracies of 0.892 and 0.850 compared to 0.841 and 0.789, respectively. Additionally, the Full model consistently surpasses the Baseline in the partial print scenarios, with the most pronounced improvement seen in heel prints---0.867, against the Baseline's 0.6751. 

Upon closer examination, the discrepancy between the accuracy of the Full model without indicator variables and that with them is minor. Across all scenarios, the largest observed difference is a mere 1.9 percentage points. This indicates that both models are virtually equivalent in their predictive accuracy. Importantly, the Full model without indicator variables offers a distinct operational benefit. In addition to its relative simplicity and stability, it forgoes the need for human intervention in classifying a test pair, dropping the requirement to label it as blurry, partial, or otherwise. This feature streamlines the process, making it fully automated. 

Tables \ref{tab:baseline_model_performance} and \ref{tab:full_model_performance} (see Appendix \ref{app:Tables}) show more details about the performance metrics of the Full model without indicator variables and the baseline model. In terms of AUC, the Full model showcases commendable consistency. While its performance on the Pristine AN scenario shows a slight dip of 0.1 percentage points from the baseline model's optimal AUC, the gains in other challenging scenarios are very pronounced. Particularly in the blurry and partial print scenarios, the Full model exhibits enhanced robustness, outperforming the baseline model's AUC by notable margins. Furthermore, the sharp uptick in AUC to 0.955 for the Pristine 150 scenario, from the baseline's 0.78, illustrates the model's superior capacity to generalise across different shoe models. These results further validate the advantage of integrating a diverse range of training scenarios for optimising model performance. 
\subsubsection{B. Full vs Category Models} \label{B}

\begin{figure*}[t]
  \centering
    \includegraphics[width=\textwidth]{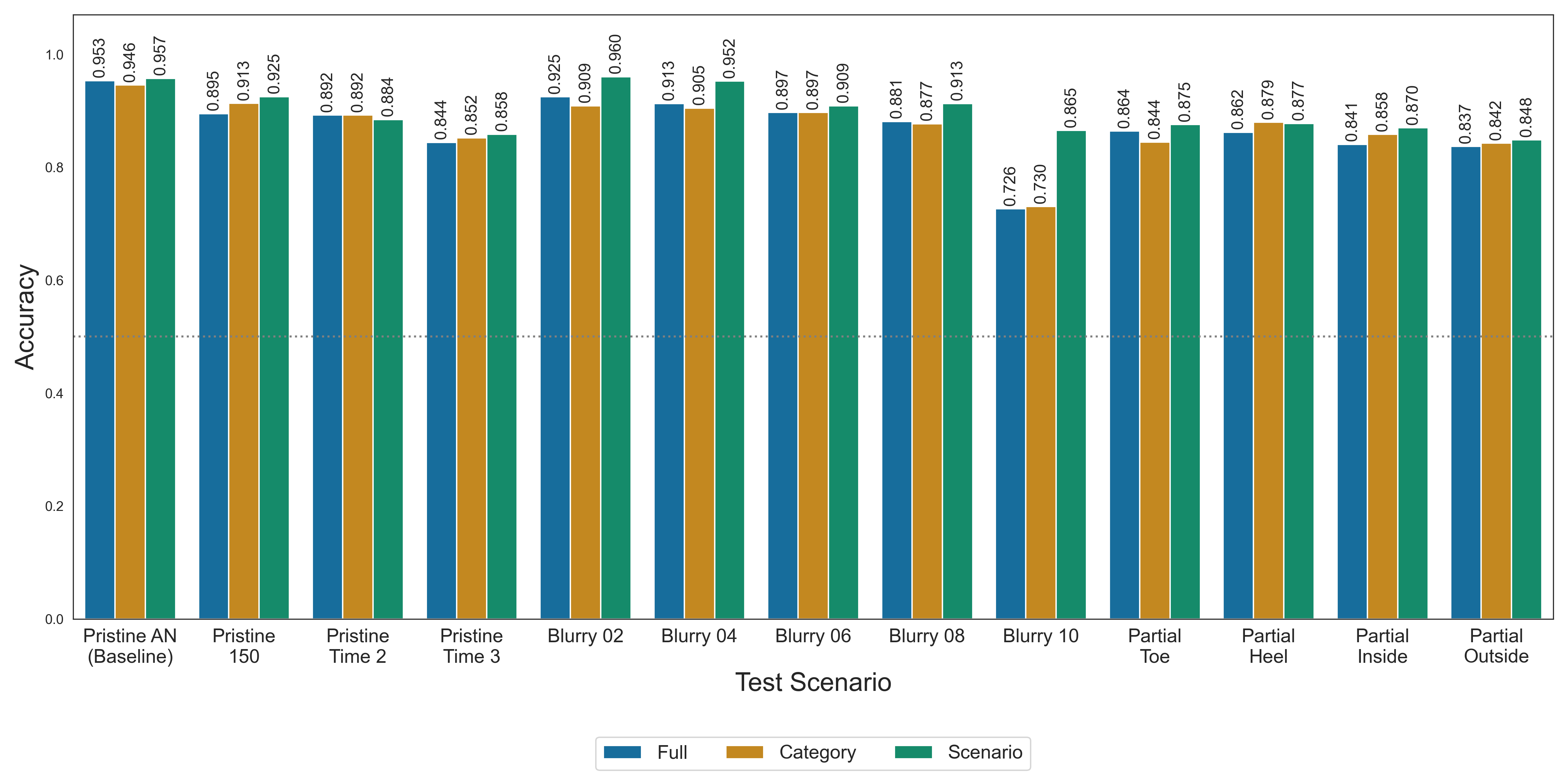}
  \caption{Comparing the test accuracy of the proposed Full model to the Category and the Scenario models. }
  \label{fig:full_model_accuracy}
\end{figure*}

Figure \ref{fig:full_model_accuracy} presents a comparative assessment of three models: the Full model, the Category models, and the Scenario models. A Category model is trained using the training datasets of each scenario within a specific category. For instance, the pristine category model is trained utilising the Pristine AN, Pristine 150, Pristine Time 2, and Pristine Time 3 scenarios. In contrast, a Scenario model is exclusively trained on the dataset pertaining to the specific scenario on which it will be tested. 

Between conventional methodologies and the ideal benchmark, we find the rationale for our Category and Scenario models. In the current landscape, Category models reflect the standard practise, where training is typically tailored to distinct conditions. For instance, models designed for blurry scenarios are trained solely on blurry datasets \citep{park2022effect}. On the other hand, the Scenario model serves as an optimal benchmark. The core idea is that, for any dataset with a defined train-test split, a model trained solely on that dataset's training set should ideally surpass models that utilise external data. This idea is anchored in the belief that a model finely calibrated to the subtleties of a specific dataset would naturally perform well in predicting outcomes on its corresponding test set. 

In a majority of scenarios, the Full model's performance closely mirrors that of the Category models. This observation suggests that training on a comprehensive dataset can yield benefits on par with training tailored to specific categories, a trend that is especially pronounced in the pristine and partial scenarios. 

It is noteworthy that the Scenario models only slightly outpace both the Full model and the Category models. This subtle difference in performance underscores the notion that broad training datasets can offer a competitive edge similar to highly specialised training approaches. However, a distinct exception can be observed in the blurry level 10 scenario. Here, both the Full and Category models witness a significant dip in performance. This underscores the inherent difficulties associated with such heightened levels of blurriness, suggesting that even tailored training strategies might not always yield optimal outcomes.  

Figure \ref{fig:full_var_imp}, Appendix \ref{app:Results} shows the variable importance of the features within the Full model. Comparing this plot with Figure \ref{fig:baseline_var_imp}, Appendix \ref{app:Results} we gain insights into the robustness of various features across distinct scenarios, pinpointing those that remain scenario-invariant. A clear example is NCC, which stands out with the greatest variable importance in both models, indicating its scenario-invariance. Further evidence of scenario-invariant features can be drawn from the consistent prominence of PSR, \(\mathcal{Q}\) proportion overlap with \(d = 10\), and \(\mathcal{K}\) proportion overlap with \(d = 10\). 

On the flip side, the comparison highlights certain features that show a diminished robustness across scenarios. For instance, all three Jaccard Index metrics, which have relatively large variable importance in the baseline model, experience a decline in the Full model relative to other metrics. Similarly, while proportion overlap metrics predominantly maintain high variable importance in the baseline model, their standing in the Full model exhibits greater variability. 

In some cases, the inverse is also true. Features such as \(\text{CPM}_{100}\) and \(\text{TWM}_{20}\), which initially possess limited importance in the baseline model, witness an elevated importance in the Full model. This suggests that these features might gain relevance in more diverse and complex modelling contexts.
\subsubsection{C. Leave-One-Out Models}

To better evaluate the generalisability of our models to new scenarios, we utilise a ``leave-one-out'' method for training and testing both the Full and Category models. This approach helps us understand how the model might perform in real-world scenarios that fall outside our predefined training data categories. In practise, professionals often encounter shoe prints that do not fit neatly into established categories. For instance, a blurry print might not strictly align with our predefined blurriness levels. Similarly, while we categorise partial prints into heel, toe, inside, and outside, in the field, a partial print might not fit neatly into these subdivisions. It is important to ensure our models can still make accurate predictions in nuanced cases. 

\begin{figure*}[t]
  \centering
    \includegraphics[width=\textwidth]{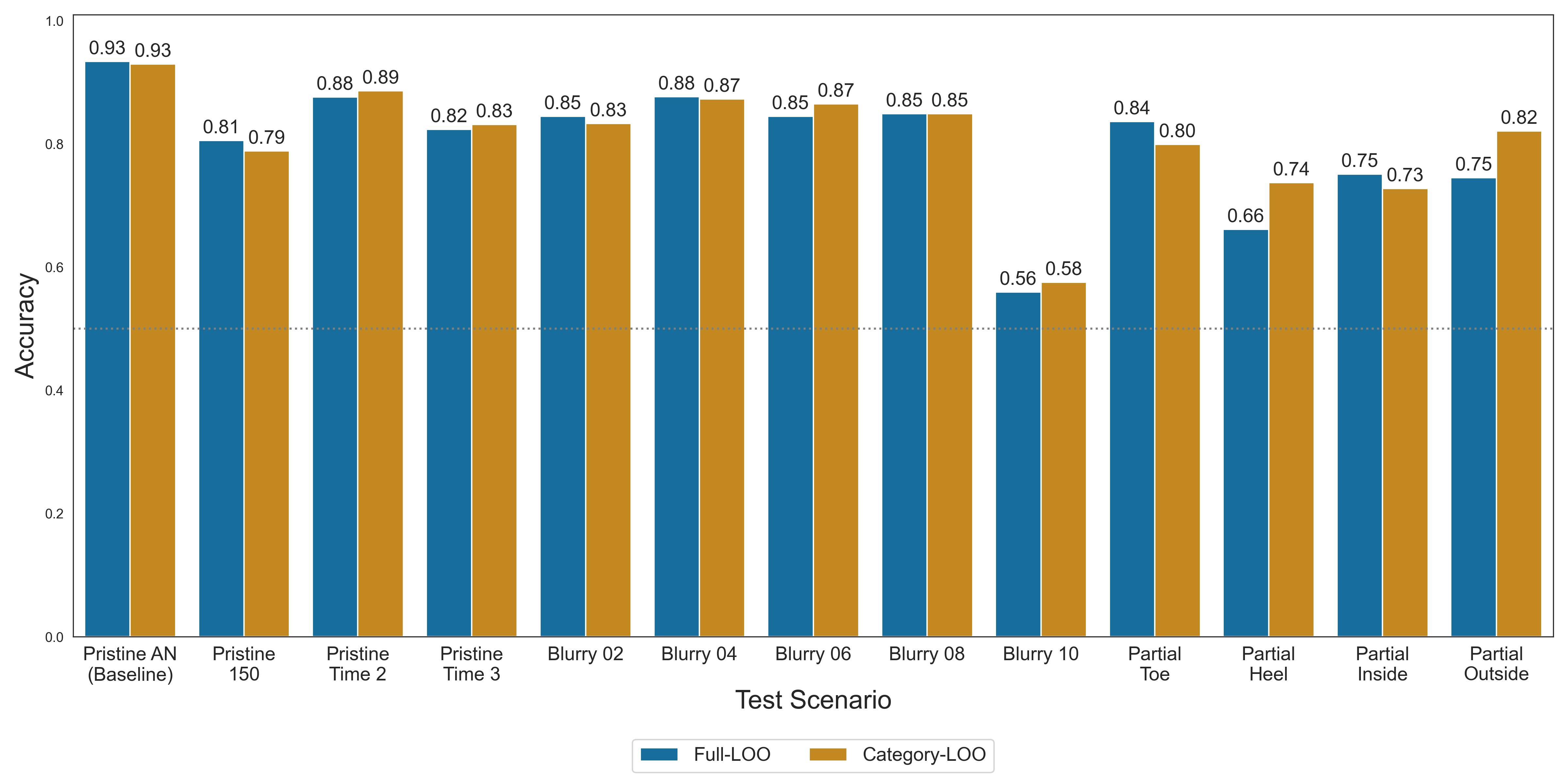}
  \caption{Comparing the test accuracy of the Full Leave-One-Out model to the Category Leave-One-Out model across each test scenario.}
  \label{fig:full_model_generalizability}
\end{figure*}

For every test scenario, our model is trained using all the available datasets, excluding the one corresponding to the current test. This excluded set is then used for testing. To compare our method with current standard practises, we apply the same leave-one-out approach to our Category models. Here, each Category model is trained on all available datasets for a given category, except for the current test scenario. The results of this methodology can be viewed in Figure \ref{fig:full_model_generalizability}. 

From Figure \ref{fig:full_model_generalizability}, it becomes apparent that the performance differences between the Full Leave-One-Out (Full-LOO) model and the Category Leave-One-Out (Category-LOO) model are relatively minimal. This is particularly evident in the pristine and blurry categories, where the largest observed difference is just 2 percentage points. For partial print scenarios, we observe modestly larger differences in performance between the models. Specifically, in the partial outside and heel prints scenarios, the deviations in accuracy reach 7 and 8 percentage points, respectively. However, no model consistently outperforms the other. Among the four partial scenarios evaluated, the Full-LOO model excels in two (partial toe and partial inside), while the Category-LOO model performs better in the remaining two (partial outside and partial heel). 

In conclusion, our exploration indicates that the Full model's ability to generalise to new and potentially unpredictable scenarios is largely on par with the Category model. This finding reaffirms the efficacy of the Full model in dealing with a wide array of shoe print scenarios and advocates for its potential value in real-world applications. \\

\section{Discussion} \label{discussion}

The primary limitation of our work lies in the nature of the data available to us. Our test accuracy and robustness checks rely on the assumption that our simulated crime scene shoeprint data reasonably replicate true crime scene data. While we exhaust the available possibilities of simulating crime scene shoeprints, the data do not quite replicate crime scene prints. For example, the blurry shoeprint data available to us was created by layering sheets of paper between the outsole of a shoe and the scanner. While these data are less pristine than the scans without paper, they do not perfectly replicate crime scene images of smudged shoeprints. 

Additionally, the scenarios we test only cover a subsection of all possible crime scene shoeprints. For instance, there is no data with which we can compare the results of running our algorithm on shoeprint outsole impressions left in different materials. Forensic researchers at CSAFE are working on creating more types of data by imaging shoeprints created with fake blood and dust. Future work may involve testing the performance of our pipeline on these new data. 

Another limitation is that our results may have been affected by the presence of noise in the shoeprint scans. As a potential result of marks on the scanner used to generate the shoeprint data, some of the shoeprints, after image processing, have clusters of points outside the main body of the shoeprint. These noise points have the ability to worsen ICP alignment as well as bias some of our similarity metrics (e.g., proportion overlap will be underestimated if there is noise in one shoeprint in a pair). We opt not to use image editing software to manually remove noise in the spirit of keeping our pipeline fully automated and free of human intervention. We considered training a neural network to create a mask around the true shoeprint to remove noise automatically, but we determined that we did not have sufficient quantities of data to do so. A potential extension of our work could consist of incorporating a masking algorithm into the image processing step of our pipeline in order to potentially further improve classification accuracy. 

Finally, one could build on our work by implementing SHAP (SHapley Additive exPlanations) value analysis into model predictions \citep{lundberg2017}. This framework would allow for increased interpretability by assigning each feature in the model an importance value for every new prediction. With this addition, examiners would be able to identify which similarity metrics are most influential in determining the classification decision. Steps taken towards increased explainability could be a pivotal step in working towards responsible, data-driven forensic science.





\clearpage
\begin{appendices}

\section{Method}
\label{app:Method}
\begin{minipage*}{\textwidth}
    \centering
    \includegraphics[width=\textwidth]{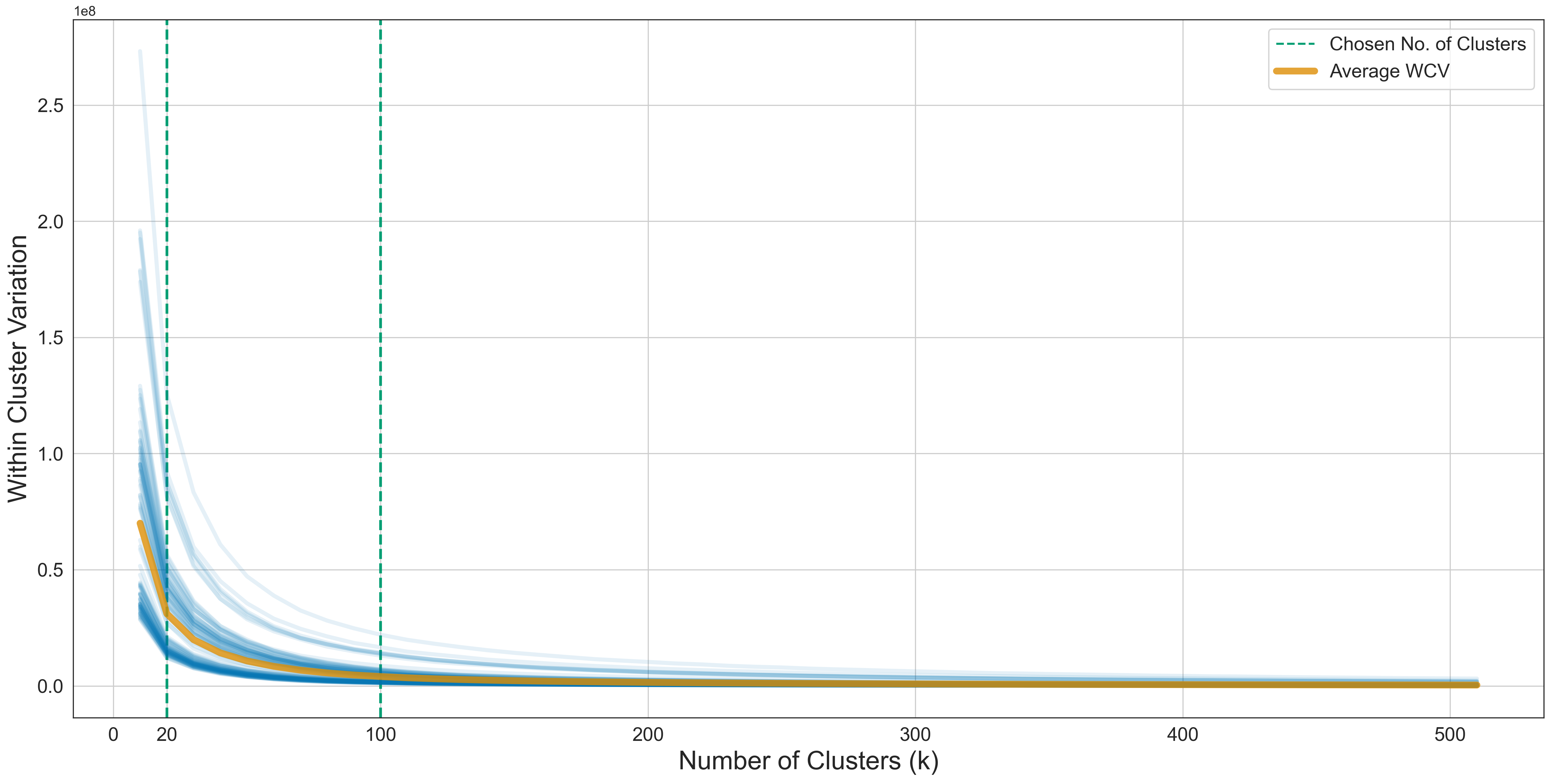}
    \captionsetup{skip=5pt}
    \captionof{figure}{Elbow Plot for selecting the optimal k using Within Cluster Variation.}
    \label{fig:cluster_elbow}
\end{minipage*}

\clearpage
\section{Result}
\label{app:Results}
\begin{minipage*}{\textwidth}
    \centering
    \includegraphics[width=\textwidth]{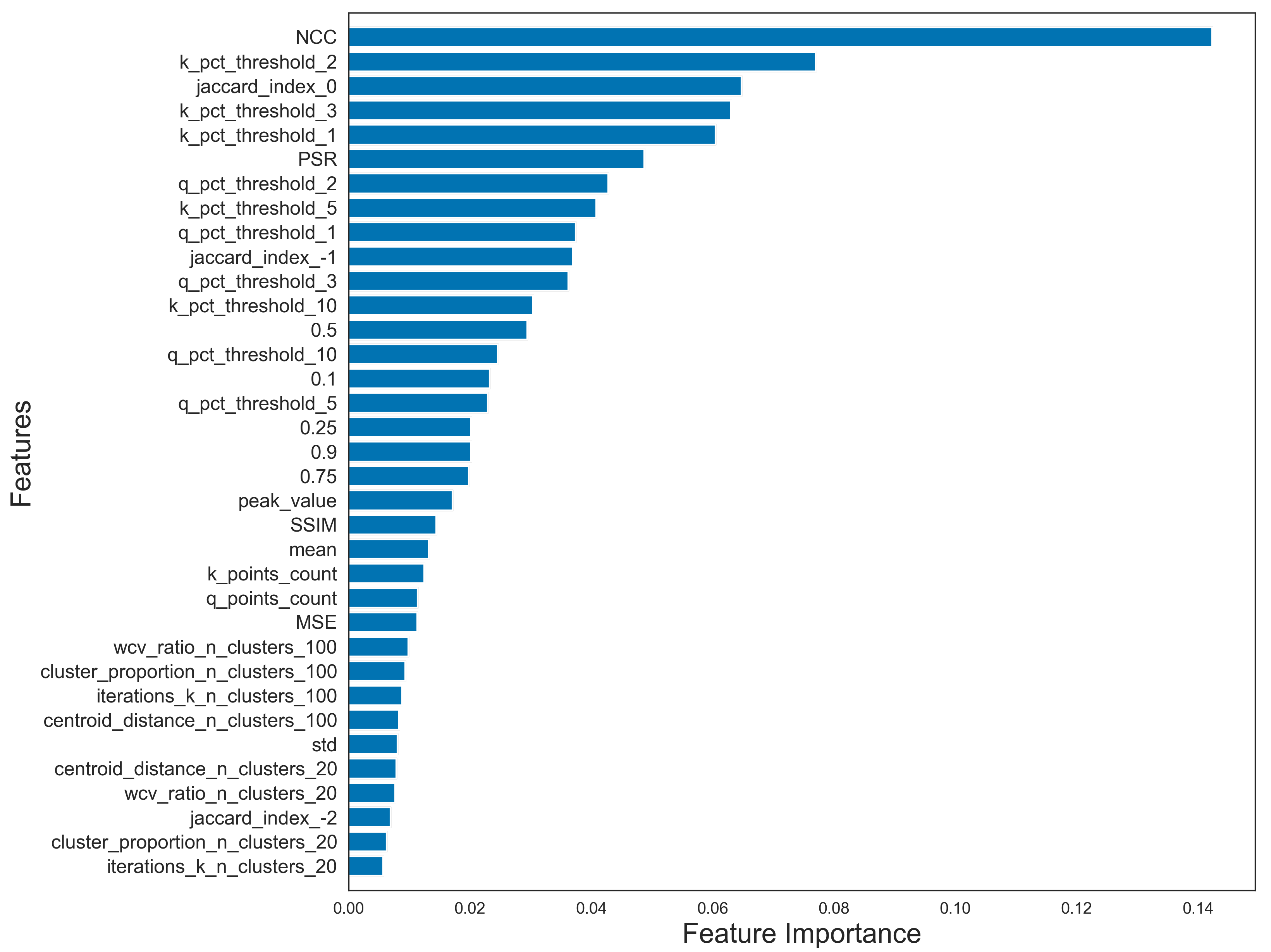}
    \captionsetup{skip=5pt}
    \captionof{figure}{Random forest variable importance for model trained on Pristine AN (baseline)}
    \label{fig:baseline_var_imp}
\end{minipage*}

\begin{figure*}[]
  \centering
    \includegraphics[width=\textwidth]{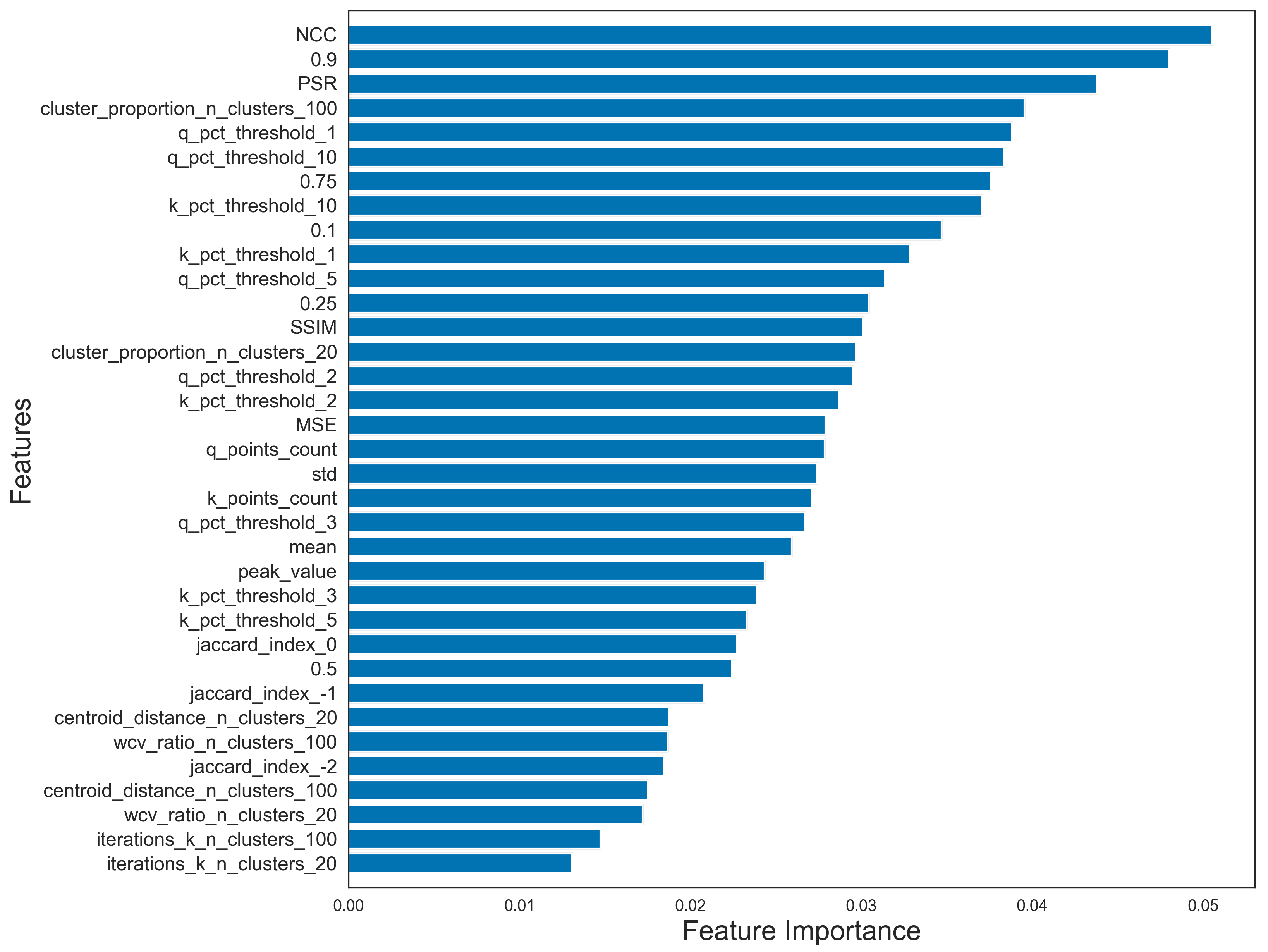}
    \caption{Random forest variable importance for Full model}
    \label{fig:full_var_imp}
\end{figure*}

\clearpage
\section{Tables}
\label{app:Tables}

\begin{minipage*}{\textwidth}
    \centering
    \begin{tabular}{lcccc}
        \toprule
        \textbf{Test Scenario} & \textbf{AUC} & \textbf{Optimal Threshold} & \textbf{Sensitivity} & \textbf{Specificity} \\
        \midrule
        Pristine AN (baseline) & 0.990 & 0.397 & 0.942 & 0.973 \\
        Pristine 150 & 0.780 & 0.441 & 0.628 & 0.788 \\
        Pristine Time 2 & 0.948 & 0.200 & 0.741 & 0.945 \\
        Pristine Time 3 & 0.911 & 0.089 & 0.607 & 0.972 \\
        Blurry 02 & 0.944 & 0.188 & 0.667 & 1.000 \\
        Blurry 04 & 0.499 & 0.156 & 0.024 & 0.976 \\
        Blurry 06 & 0.284 & 0.150 & 0.000 & 0.992 \\
        Blurry 08 & 0.261 & 0.151 & 0.000 & 0.992 \\
        Blurry 10 & 0.334 & 0.154 & 0.000 & 0.968 \\
        Partial Toe & 0.905 & 0.358 & 0.720 & 0.903 \\
        Partial Heel & 0.841 & 0.772 & 0.821 & 0.529 \\
        Partial Inside & 0.854 & 0.406 & 0.689 & 0.829 \\
        Partial Outside & 0.841 & 0.542 & 0.743 & 0.716 \\
        \bottomrule
    \end{tabular}
    \captionsetup{skip=5pt}
    \captionof{table}{Baseline model performance metrics.}
    \label{tab:baseline_model_performance}
\end{minipage*}

\vspace{0.3in}
\begin{minipage*}{\textwidth}
    \centering
    \begin{tabular}{lcccc}
        \toprule
        \textbf{Test Scenario} & \textbf{AUC} & \textbf{Optimal Threshold} & \textbf{Sensitivity} & \textbf{Specificity} \\
        \midrule
        Pristine AN (baseline) & 0.989 & 0.622 & 0.977 & 0.930 \\
        Pristine 150 & 0.955 & 0.380 & 0.830 & 0.960 \\
        Pristine Time 2 & 0.954 & 0.500 & 0.890 & 0.894 \\
        Pristine Time 3 & 0.937 & 0.354 & 0.741 & 0.947 \\
        Blurry 02 & 0.971 & 0.728 & 0.952 & 0.897 \\
        Blurry 04 & 0.972 & 0.546 & 0.937 & 0.889 \\
        Blurry 06 & 0.959 & 0.521 & 0.897 & 0.897 \\
        Blurry 08 & 0.969 & 0.327 & 0.778 & 0.984 \\
        Blurry 10 & 0.908 & 0.232 & 0.476 & 0.976 \\
        Partial Toe & 0.934 & 0.404 & 0.778 & 0.949 \\
        Partial Heel & 0.942 & 0.508 & 0.868 & 0.856 \\
        Partial Inside & 0.921 & 0.397 & 0.770 & 0.911 \\
        Partial Outside & 0.901 & 0.441 & 0.778 & 0.895 \\
        \bottomrule
    \end{tabular}
    \captionsetup{skip=5pt}
    \captionof{table}{Full model Performance metrics.}
    \label{tab:full_model_performance}
\end{minipage*}

\clearpage
\begin{sidewaystable*}
    \begin{tabular}{llllllllllllll}
    \toprule
    Similarity Metric & {\makecell[t]{Pristine\\ AN}} & {\makecell[t]{Blurry\\ 02}} & {\makecell[t]{Blurry\\ 04}} & {\makecell[t]{Blurry\\ 06}} & {\makecell[t]{Blurry\\ 08}} & {\makecell[t]{Blurry\\ 10}} & {\makecell[t]{Pristine\\ Time 2}} & {\makecell[t]{Pristine\\ Time 3}} & {\makecell[t]{Partial\\ Toe}} & {\makecell[t]{Partial\\ Heel}} & {\makecell[t]{Partial\\ Inside}} & {\makecell[t]{Partial\\ Outside}} & {\makecell[t]{Pristine\\ 150}} \\
    \midrule
    q\_points\_count & 0.0 & 1.7 & 1.74 & 1.75 & 1.73 & 1.71 & 0.04 & 0.03 & 1.43 & 1.58 & 1.63 & 1.36 & 0.81 \\
    k\_points\_count & 0.0 & 1.12 & 1.12 & 1.12 & 1.12 & 1.12 & 0.09 & 0.12 & 1.41 & 1.56 & 1.57 & 1.32 & 0.82 \\
    mean & 0.0 & 0.46 & 0.66 & 0.67 & 0.83 & 0.88 & 0.27 & 0.28 & 0.27 & 0.2 & 0.27 & 0.21 & 0.26 \\
    std & 0.0 & 0.4 & 0.52 & 0.46 & 0.58 & 0.55 & 0.21 & 0.19 & 0.12 & 0.35 & 0.09 & 0.34 & 0.32 \\
    0.1 & 0.0 & 0.78 & 1.54 & 1.65 & 1.69 & 1.75 & 0.65 & 0.89 & 0.65 & 0.45 & 0.66 & 0.51 & 0.55 \\
    0.25 & 0.0 & 0.88 & 1.64 & 1.72 & 1.77 & 1.85 & 0.64 & 0.87 & 0.61 & 0.48 & 0.66 & 0.52 & 0.56 \\
    0.5 & 0.0 & 0.93 & 1.61 & 1.73 & 1.8 & 1.87 & 0.63 & 0.84 & 0.54 & 0.47 & 0.63 & 0.49 & 0.52 \\
    0.75 & 0.0 & 0.52 & 1.13 & 1.33 & 1.54 & 1.68 & 0.45 & 0.57 & 0.53 & 0.42 & 0.56 & 0.52 & 0.41 \\
    0.9 & 0.0 & 0.4 & 0.52 & 0.56 & 0.68 & 0.77 & 0.19 & 0.19 & 0.17 & 0.14 & 0.16 & 0.24 & 0.22 \\
    centroid\_distance\_n\_clusters\_20 & 0.0 & 0.28 & 0.7 & 1.02 & 1.07 & 1.14 & 0.24 & 0.39 & 0.29 & 0.64 & 0.55 & 0.52 & 0.12 \\
    cluster\_proportion\_n\_clusters\_20 & 0.0 & 1.39 & 1.58 & 1.74 & 1.78 & 1.83 & 0.21 & 0.4 & 0.38 & 0.12 & 0.5 & 0.52 & 0.25 \\
    iterations\_k\_n\_clusters\_20 & 0.0 & 0.17 & 0.26 & 0.37 & 0.42 & 0.47 & 0.15 & 0.14 & 0.16 & 0.13 & 0.64 & 0.37 & 0.18 \\
    wcv\_ratio\_n\_clusters\_20 & 0.0 & 0.35 & 0.32 & 0.44 & 0.45 & 0.5 & 0.15 & 0.16 & 0.23 & 0.22 & 0.36 & 0.39 & 0.26 \\
    centroid\_distance\_n\_clusters\_100 & 0.0 & 0.12 & 0.15 & 0.34 & 0.25 & 0.28 & 0.16 & 0.13 & 0.18 & 0.19 & 0.48 & 0.07 & 0.1 \\
    cluster\_proportion\_n\_clusters\_100 & 0.0 & 1.31 & 1.59 & 1.7 & 1.72 & 1.75 & 0.22 & 0.37 & 0.5 & 0.2 & 0.5 & 0.35 & 0.21 \\
    iterations\_k\_n\_clusters\_100 & 0.0 & 0.38 & 0.3 & 0.33 & 0.31 & 0.28 & 0.12 & 0.07 & 0.14 & 0.18 & 0.6 & 0.14 & 0.25 \\
    wcv\_ratio\_n\_clusters\_100 & 0.0 & 0.51 & 0.57 & 0.77 & 0.68 & 0.69 & 0.1 & 0.11 & 0.29 & 0.09 & 0.33 & 0.22 & 0.43 \\
    q\_pct\_threshold\_1 & 0.0 & 0.97 & 1.49 & 1.56 & 1.6 & 1.66 & 0.67 & 0.94 & 0.52 & 0.48 & 0.63 & 0.38 & 0.56 \\
    k\_pct\_threshold\_1 & 0.0 & 1.7 & 1.84 & 1.85 & 1.85 & 1.85 & 0.64 & 0.85 & 0.62 & 0.4 & 0.68 & 0.42 & 0.56 \\
    q\_pct\_threshold\_2 & 0.0 & 1 & 1.52 & 1.61 & 1.65 & 1.71 & 0.67 & 0.95 & 0.54 & 0.49 & 0.66 & 0.38 & 0.58 \\
    k\_pct\_threshold\_2 & 0.0 & 1.69 & 1.85 & 1.87 & 1.87 & 1.87 & 0.64 & 0.85 & 0.65 & 0.39 & 0.71 & 0.44 & 0.57 \\
    q\_pct\_threshold\_3 & 0.0 & 0.96 & 1.54 & 1.65 & 1.71 & 1.77 & 0.67 & 0.93 & 0.56 & 0.5 & 0.69 & 0.42 & 0.6 \\
    k\_pct\_threshold\_3 & 0.0 & 1.65 & 1.85 & 1.89 & 1.89 & 1.89 & 0.64 & 0.85 & 0.68 & 0.37 & 0.73 & 0.46 & 0.59 \\
    q\_pct\_threshold\_5 & 0.0 & 0.62 & 1.39 & 1.6 & 1.72 & 1.81 & 0.6 & 0.83 & 0.58 & 0.48 & 0.67 & 0.48 & 0.6 \\
    k\_pct\_threshold\_5 & 0.0 & 1.45 & 1.79 & 1.86 & 1.87 & 1.88 & 0.57 & 0.77 & 0.71 & 0.29 & 0.72 & 0.51 & 0.59 \\
    q\_pct\_threshold\_10 & 0.0 & 0.37 & 0.65 & 0.89 & 1.29 & 1.59 & 0.36 & 0.48 & 0.5 & 0.35 & 0.49 & 0.48 & 0.46 \\
    k\_pct\_threshold\_10 & 0.0 & 0.66 & 1.21 & 1.43 & 1.45 & 1.51 & 0.31 & 0.46 & 0.62 & 0.21 & 0.53 & 0.52 & 0.46 \\
    peak\_value & 0.0 & 1.56 & 1.65 & 1.66 & 1.66 & 1.65 & 0.38 & 0.58 & 1.24 & 1.25 & 1.47 & 1.09 & 0.55 \\
    MSE & 0.0 & 1.4 & 1.42 & 1.41 & 1.4 & 1.4 & 0.04 & 0.07 & 0.2 & 1.51 & 1.33 & 0.55 & 0.97 \\
    SSIM & 0.0 & 0.93 & 0.86 & 0.78 & 0.72 & 0.67 & 0.11 & 0.13 & 0.13 & 1.4 & 1.13 & 0.54 & 0.68 \\
    NCC & 0.0 & 1.46 & 1.72 & 1.75 & 1.76 & 1.75 & 0.76 & 1 & 0.31 & 0.73 & 0.31 & 0.24 & 0.38 \\
    PSR & 0.0 & 0.43 & 0.73 & 0.84 & 0.91 & 0.92 & 0.58 & 0.72 & 0.23 & 1.45 & 1.1 & 0.52 & 0.83 \\
    jaccard\_index\_0 & 0.0 & 1.47 & 1.7 & 1.73 & 1.73 & 1.73 & 0.67 & 0.9 & 0.54 & 0.44 & 0.59 & 0.38 & 0.5 \\
    jaccard\_index\_-1 & 0.0 & 1.22 & 1.66 & 1.77 & 1.82 & 1.87 & 0.64 & 0.87 & 0.67 & 0.55 & 0.84 & 0.44 & 0.81 \\
    jaccard\_index\_-2 & 0.0 & 0.57 & 0.75 & 0.89 & 1.01 & 1.15 & 0.31 & 0.59 & 0.31 & 0.2 & 0.36 & 0.36 & 0.84 \\
    \bottomrule
    \end{tabular}
    \caption{Earth Mover's Distance, for each similarity metric, between each scenario and the Pristine AN (baseline) among known mated pairs. Similarity metrics are standardised using its mean and \\ standard deviation calculated from the combined dataset (scenario combined with baseline).}
    \label{tab:earth_movers_distance_mated}
\end{sidewaystable*}

\clearpage
\begin{sidewaystable*}
    \begin{tabular}{llllllllllllll}
    \toprule
    Similarity Metric & {\makecell[t]{Pristine\\ AN}} & {\makecell[t]{Blurry\\ 02}} & {\makecell[t]{Blurry\\ 04}} & {\makecell[t]{Blurry\\ 06}} & {\makecell[t]{Blurry\\ 08}} & {\makecell[t]{Blurry\\ 10}} & {\makecell[t]{Pristine\\ Time 2}} & {\makecell[t]{Pristine\\ Time 3}} & {\makecell[t]{Partial\\ Toe}} & {\makecell[t]{Partial\\ Heel}} & {\makecell[t]{Partial\\ Inside}} & {\makecell[t]{Partial\\ Outside}} & {\makecell[t]{Pristine\\ 150}} \\
    \midrule
    q\_points\_count & 0.0 & 1.73 & 1.77 & 1.78 & 1.76 & 1.74 & 0.2 & 0.35 & 1.47 & 1.62 & 1.66 & 1.39 & 0.81 \\
    k\_points\_count & 0.0 & 1.16 & 1.16 & 1.16 & 1.16 & 1.16 & 0.16 & 0.44 & 1.48 & 1.57 & 1.57 & 1.38 & 0.79 \\
    mean & 0.0 & 0.46 & 0.4 & 0.48 & 0.52 & 0.47 & 0.12 & 0.19 & 0.18 & 0.31 & 0.42 & 0.21 & 0.79 \\
    std & 0.0 & 0.46 & 0.46 & 0.48 & 0.5 & 0.44 & 0.1 & 0.18 & 0.19 & 0.43 & 0.12 & 0.5 & 0.99 \\
    0.1 & 0.0 & 0.87 & 1.15 & 1.23 & 1.37 & 1.37 & 0.13 & 0.33 & 0.5 & 0.47 & 0.85 & 0.38 & 1.14 \\
    0.25 & 0.0 & 1.02 & 1.24 & 1.33 & 1.46 & 1.46 & 0.1 & 0.37 & 0.52 & 0.48 & 0.86 & 0.35 & 1.16 \\
    0.5 & 0.0 & 0.84 & 0.98 & 1.21 & 1.37 & 1.22 & 0.12 & 0.34 & 0.53 & 0.48 & 0.75 & 0.4 & 1.1 \\
    0.75 & 0.0 & 0.39 & 0.56 & 0.78 & 1.01 & 1.1 & 0.15 & 0.21 & 0.5 & 0.45 & 0.65 & 0.44 & 0.86 \\
    0.9 & 0.0 & 0.41 & 0.37 & 0.47 & 0.5 & 0.52 & 0.11 & 0.13 & 0.17 & 0.27 & 0.32 & 0.22 & 0.43 \\
    centroid\_distance\_n\_clusters\_20 & 0.0 & 0.2 & 0.48 & 0.83 & 0.85 & 0.9 & 0.11 & 0.11 & 0.29 & 0.55 & 0.45 & 0.66 & 0.49 \\
    cluster\_proportion\_n\_clusters\_20 & 0.0 & 1.31 & 1.51 & 1.68 & 1.72 & 1.82 & 0.09 & 0.16 & 0.6 & 0.17 & 0.83 & 0.49 & 0.65 \\
    iterations\_k\_n\_clusters\_20 & 0.0 & 0.19 & 0.26 & 0.36 & 0.36 & 0.41 & 0.11 & 0.11 & 0.29 & 0.14 & 0.42 & 0.29 & 0.13 \\
    wcv\_ratio\_n\_clusters\_20 & 0.0 & 0.29 & 0.31 & 0.28 & 0.41 & 0.56 & 0.07 & 0.11 & 0.27 & 0.35 & 0.6 & 0.48 & 0.27 \\
    centroid\_distance\_n\_clusters\_100 & 0.0 & 0.12 & 0.14 & 0.22 & 0.13 & 0.13 & 0.17 & 0.07 & 0.24 & 0.23 & 0.5 & 0.16 & 0.72 \\
    cluster\_proportion\_n\_clusters\_100 & 0.0 & 1.37 & 1.62 & 1.76 & 1.77 & 1.81 & 0.09 & 0.42 & 0.72 & 0.14 & 0.91 & 0.36 & 0.71 \\
    iterations\_k\_n\_clusters\_100 & 0.0 & 0.37 & 0.31 & 0.39 & 0.45 & 0.36 & 0.16 & 0.23 & 0.14 & 0.18 & 0.39 & 0.26 & 0.41 \\
    wcv\_ratio\_n\_clusters\_100 & 0.0 & 0.49 & 0.58 & 0.73 & 0.67 & 0.74 & 0.17 & 0.25 & 0.31 & 0.28 & 0.49 & 0.27 & 0.46 \\
    q\_pct\_threshold\_1 & 0.0 & 0.96 & 1.18 & 1.23 & 1.36 & 1.42 & 0.1 & 0.36 & 0.42 & 0.56 & 0.97 & 0.45 & 1.4 \\
    k\_pct\_threshold\_1 & 0.0 & 1.82 & 1.89 & 1.9 & 1.9 & 1.89 & 0.07 & 0.18 & 0.52 & 0.48 & 1.06 & 0.32 & 1.42 \\
    q\_pct\_threshold\_2 & 0.0 & 0.98 & 1.2 & 1.27 & 1.39 & 1.46 & 0.1 & 0.38 & 0.45 & 0.58 & 1 & 0.47 & 1.42 \\
    k\_pct\_threshold\_2 & 0.0 & 1.77 & 1.86 & 1.87 & 1.87 & 1.86 & 0.08 & 0.2 & 0.55 & 0.49 & 1.09 & 0.34 & 1.43 \\
    q\_pct\_threshold\_3 & 0.0 & 0.95 & 1.17 & 1.26 & 1.4 & 1.48 & 0.1 & 0.37 & 0.49 & 0.59 & 1.03 & 0.48 & 1.38 \\
    k\_pct\_threshold\_3 & 0.0 & 1.68 & 1.8 & 1.82 & 1.82 & 1.81 & 0.09 & 0.23 & 0.6 & 0.5 & 1.11 & 0.35 & 1.39 \\
    q\_pct\_threshold\_5 & 0.0 & 0.71 & 1 & 1.14 & 1.32 & 1.43 & 0.11 & 0.31 & 0.53 & 0.61 & 1.01 & 0.49 & 1.27 \\
    k\_pct\_threshold\_5 & 0.0 & 1.44 & 1.62 & 1.69 & 1.68 & 1.66 & 0.1 & 0.26 & 0.64 & 0.51 & 1.1 & 0.35 & 1.29 \\
    q\_pct\_threshold\_10 & 0.0 & 0.37 & 0.66 & 0.89 & 1.16 & 1.3 & 0.15 & 0.21 & 0.48 & 0.55 & 0.79 & 0.47 & 1 \\
    k\_pct\_threshold\_10 & 0.0 & 0.67 & 0.95 & 1.18 & 1.15 & 1.12 & 0.09 & 0.23 & 0.61 & 0.39 & 0.93 & 0.32 & 1.06 \\
    peak\_value & 0.0 & 1.66 & 1.73 & 1.73 & 1.73 & 1.72 & 0.12 & 0.4 & 1.35 & 1.4 & 1.6 & 1.08 & 0.98 \\
    MSE & 0.0 & 1.46 & 1.49 & 1.48 & 1.47 & 1.47 & 0.19 & 0.43 & 0.22 & 1.55 & 1.33 & 0.58 & 0.85 \\
    SSIM & 0.0 & 0.97 & 0.96 & 0.92 & 0.87 & 0.83 & 0.19 & 0.36 & 0.16 & 1.43 & 1.09 & 0.58 & 0.48 \\
    NCC & 0.0 & 1.48 & 1.73 & 1.73 & 1.74 & 1.74 & 0.17 & 0.13 & 0.29 & 1 & 0.44 & 0.7 & 1.21 \\
    PSR & 0.0 & 0.67 & 0.34 & 0.33 & 0.36 & 0.39 & 0.24 & 0.3 & 0.28 & 1.6 & 1.14 & 0.86 & 0.45 \\
    jaccard\_index\_0 & 0.0 & 1.67 & 1.77 & 1.78 & 1.79 & 1.79 & 0.11 & 0.25 & 0.49 & 0.52 & 0.96 & 0.39 & 1.28 \\
    jaccard\_index\_-1 & 0.0 & 1.2 & 1.46 & 1.57 & 1.63 & 1.66 & 0.08 & 0.34 & 0.68 & 0.73 & 1.19 & 0.59 & 1.27 \\
    jaccard\_index\_-2 & 0.0 & 0.63 & 0.58 & 0.83 & 0.78 & 0.82 & 0.1 & 0.47 & 0.27 & 0.32 & 0.54 & 0.41 & 1.22 \\
    \bottomrule
    \end{tabular}
    \caption{Earth Mover's Distance, for each similarity metric, between each scenario and the Pristine AN (baseline) among known non-mated pairs. Similarity metrics are standardised similar to Table \ref{tab:earth_movers_distance_mated}.}
    \label{tab:earth_movers_distance_non_mated}
\end{sidewaystable*}

\end{appendices}

\clearpage
\section{Competing interests}
No competing interest is declared.

\section{Author contributions statement}

D.J., S.K., L.L., Y.W., and A.Z. contributed equally in conducting the research, analyzing the results, and writing the manuscript. X.C., A.P., and E.U. advised the research and provided suggestions, support, and outreach communication. All authors,  D.J., S.K., L.L., Y.W., A.Z., X.C., A.P., and E.U., reviewed the manuscript.


\section{Acknowledgments}
We would like to thank Professor Alicia Carriquiry, Hana Lee, Gautham Venkatasubramanian, and the rest of the team at CSAFE for making the footwear data they collect accessible to us as well as for providing feedback on our work. 

We acknowledge the work of Simon Angoluan at Williams College who also participated in conducting the research and analyzing the results. 

This work was funded by Williams College Science Center and by the National Science Foundation (NSF) via the Williams College SMALL Undergraduate Research Project through grants DMS2241623 and DMS1947438. 

This work was partially funded by the Center for Statistics and Applications in Forensic Evidence (CSAFE) through Cooperative Agreements 70NANB15H176 and 70NANB20H019 between NIST and Iowa State University, which includes activities carried out at Carnegie Mellon University, Duke University, University of California Irvine, University of Virginia, West Virginia University, University of Pennsylvania, Swarthmore College and University of Nebraska, Lincoln.

\bibliographystyle{plain}
\bibliography{reference}




\end{document}